\documentclass[preprint,aps,showpacs,amsmath,amsfonts,epsf,psfig,eqsecnum]{revtex4}
\usepackage{epsfig}
\draft

\newcommand{\be}{\begin{equation}}
\newcommand{\ee}{\end{equation}}
\newcommand{\bea}{\begin{eqnarray}}
\newcommand{\eea}{\end{eqnarray}}
\newcommand{\bc}{\begin{center}}
\newcommand{\ec}{\end{center}}

\newcommand{\PD}{{\partial}}
\newcommand{\btu}{\bigtriangleup}
\newcommand{\btd}{\bigtriangledown}

\newcommand{\vect}[1]{\mbox{\boldmath${#1}$}}
\renewcommand{\vec}[1]{\mbox{\boldmath${#1}$}}

\newcommand{\pr}{\prime}
\newcommand{\eps}{\varepsilon}

\newcommand{\lla}{\langle}
\newcommand{\gra}{\rangle}
\newcommand{\half}{\frac{1}{2}}
\def \a {\alpha}
\def \b {\beta}
\def \c {\chi}
\def \f{\phi}

\def \s {\sigma}
\def \g {\gamma}
\def \h {\eta}
\def \r {\rho}
\def \l {\lambda}
\def \k {\kappa}
\def \ep{\varepsilon}
\def \ve{\varepsilon}
\def \D {\mbox{D}}

\def \be {\begin{equation}}
\def \ee{\end{equation}}
\def \bea{\begin{eqnarray}}
\def \eea{\end{eqnarray}}
\def \m {\mu}
\def \n {\nu}
\def \t {\tau}
\def \z {\zeta}

\def\ten #1{\bf#1}
\def\tp{{\ten p}}               
\def\tv{{\ten v}}               
\def\tx{{\ten x}}
\def\tpt{{\tp_\perp}}
\def\txt{{\tx_\perp}}
\def\zt{{z_\perp}}

\def\mt{{m_\perp}}
\def\Et{{E_\perp}}

\def\ats{{\alpha_\perp^2}}
\def\zts{{z_\perp^2}}
\def\mts{{m_\perp^2}}
\def\pts{{p_\perp^2}}

\newcommand\ca{{\cal A}}
\newcommand\cb{{\cal B}}
\newcommand\cc{{\cal C}}
\newcommand\cf{{\cal F}}

\newcommand\ci{{\cal I}}
\newcommand\cj{{\cal J}}
\newcommand\ck{{\cal K}}
\newcommand\cl{{\cal L}}
\newcommand\cq{{\cal Q}}

\include{scrload}
\newcommand\Lra{\Longrightarrow}

\newcommand\dd{\mbox{d}}   
\renewcommand\exp{\mbox{\rm exp}}  
\newcommand\eqq{{\rm eq}}       
\def\sks{,\;\;\;\;}

\newfam\scrfam                                          
\global\font\twelvescr=rsfs10 scaled\magstep1%
\global\font\eightscr=rsfs7 scaled\magstep1%
\global\font\sixscr=rsfs5 scaled\magstep1%
\skewchar\twelvescr='177\skewchar\eightscr='177\skewchar\sixscr='177%
\textfont\scrfam=\twelvescr\scriptfont\scrfam=\eightscr
\scriptscriptfont\scrfam=\sixscr
\begin{document}

\title{ Relativistic Dynamics of Non-ideal Fluids: Viscous and heat-conducting
fluids \\
        II. Transport properties and microscopic description of relativistic
	nuclear matter}
\author{Azwinndini Muronga$^{1,2}$}

\address{$^1$Centre for Theoretical Physics \& Astrophysics, 
Department of Physics, University of Cape Town, Rondebosch 7701, South Africa\\
$^2$UCT--CERN Research Centre,  
Department of Physics, University of Cape Town, Rondebosch 7701, South Africa}

\date{\today}

\begin{abstract}

In the causal theory of relativistic dissipative fluid dynamics, there are
conditions on the equation of state and other thermodynamic properties such as
the second-order coefficients of a fluid that need to be satisfied to 
guarantee that the fluid perturbations propagate causally and obey hyperbolic
equations. The second-order coefficients in the causal theory, which are the
relaxation times for the dissipative degrees of freedom and coupling constants
between different forms of dissipation (relaxation lengths), are presented for
partonic and hadronic systems. These coefficients involves  relativistic
thermodynamic integrals. The integrals are presented for general case and also
for different regimes in the temperature--chemical potential plane.  It is
shown that for a given equation of state these second-order coefficients are
not additional parameters but they are determined by the equation of state.  We
also present the prescription on the calculation of the freeze-out particle
spectra from the dynamics of relativistic non-ideal fluids. 

\end{abstract}

\maketitle

\renewcommand{\thefootnote}{\arabic{footnote}}
\setcounter{footnote}{0}
%
%
\section{Introduction}
%
Transport properties of relativistic hot and dense matter are important in
many  physical situations such as in nuclear physics, astrophysics and
cosmology. The hot and dense relativistic matter is produced in high energy
relativistic nuclear reactions such as those in heavy-ion experiments while the
hot matter is believed to have existed in the early universe and the dense
matter might be found in the quark stars and neutron stars. 
The transport properties of a system out of equilibrium are governed by 
transport coefficients which relate flows to thermodynamic forces. These
coefficients characterize the magnitude of the response of the system (flow) to
a certain disturbance (thermodynamic force). 

The space-time evolution of the produced matter in high energy nuclear 
collisions can be studied using relativistic fluid dynamics. In order to take
into account non-equilibrium (dissipative) effects we need to use relativistic
non-ideal fluid dynamics. We consider the causal theory of relativistic fluids
as presented in paper I \cite{AMI}.

In the high energy heavy ion collisions the rapid evolution of the fluid is 
governed by dissipative effects via transport coefficients such as the shear
and bulk viscosities, the thermal conductivity, the diffusion coefficients,
etc.  These transport coefficients are generally calculated via the use of
kinetic theory and thereby imply the knowledge of a collision term. The
collision term must be a representation of the hot and dense matter at hand.

In addition to the standard transport coefficients the causal  theory contains
other thermodynamic functions, namely the second-order coefficients. These
coefficients, in combination with the standard transport coefficients, are
related to the relaxation times and relaxation length  for various dissipative
processes. The relaxation length results from the  coupling between the heat
flux and viscous processes. These second-order coefficients are given by
complicated thermodynamic integrals. In this work we will present the
second-order transport coefficients in detail for future applications in the
description of the relativistic dynamics of heat-conducting and viscous matter
produced in high energy nuclear collisions.

In addition to the knowledge of the transport coefficients and second-order
thermodynamic function we need to know how to determine the form of the
equation of state. Knowledge of the equation of state is needed for analyzing
many important physical situations such as supernova explosions and neutron
star formation, high energy nuclear collisions, and the study of quark-hadron
phase transition in the early universe. Recent developments in heavy ion
collision experiments can reveal the form of the equation of state of the
nuclear matter. The physics involved in such collisions is highly complex and
it is highly nontrivial matter to extract equation of state information from
such a complex dynamical situation.

In this paper we would like to investigate the influence of the equation of
state on the transport properties of the relativistic hot and dense matter. The
equation of state in the hadronic phase is derived from the Walecka model of
nuclear matter and in the QGP phase we use the MIT Bag model. In the mixed
phase we employ the Gibbs construction for the phase equilibria.

The rest of the paper is organized as follows: In Section \ref{sec:transport}
we present the moments of relativistic Boltzmann equation in terms of the
relativistic thermodynamic integrals. In Section \ref{sec:14-fields} we present
the 14-Field Theory from the microscopic considerations. We consider small
deviations from equilibrium and employ the Grad's 14 moment method. In Section
\ref{sec:2nd} we present the second order entropy 4-current in terms of the
relaxation and coupling coefficients which are collectively referred to as the
second order coefficients. In Section \ref{sec:EoS} we present the EoS
considered in this work. In Section \ref{sec:freeze} we present the freeze-out
prescription which takes into account the dissipative corrections. In Appendix
\ref{sec:momint} we present the relativistic thermodynamic integrals. In
Appendix \ref{sec:limits} we present the various limiting cases of the
thermodynamic integrals. In Appendix \ref{sec:ultra-rel} we present the
ultra-relativistic thermodynamic integrals. In Appendix \ref{sec:T=0} we 
present the relativistic thermodynamic integrals at zero temperature.

Our units are $\hbar = c = k_{\rm B} = 1$.
The metric tensor is $g^{\mu \nu} = \mbox{diag} (+,-,-,-)$.
The scalar product of two 4-vectors $a^\mu, \,\, b^\mu$ is denoted by 
$a^\mu\, g_{\mu \nu}\, b^\nu \equiv a^\mu \, b_\mu$, and
the scalar product of two 3-vectors $\vect{a}$ and  $\vect{b}$ 
by $\vect{a} \cdot \vect{b}$.
The notations $A^{(\a\b)} \equiv 
(A^{\a\b}+A^{\b\a})/2$ and $A^{[\a\b]} \equiv  (A^{\a\b}-A^{\b\a})/2$ denote
symmetrization and antisymmetrization, respectively. The notation
$A^{\lla\a\b\gra} = [\btu_\m^{(\a}\btu_\n^{\b)}-1/3
\btu^{\a\b}\btu_{\m\n}]A^{\m\n}$ denotes the trace-free part of $A^{\m\n}$. The 4-derivative is
denoted by $\PD_\a \equiv \PD/\PD x^\a$. 
An overdot denotes $\dot{A}=u^\l \PD_\l A$. 

\section{The Relativistic Transport Equation and the Moment Equations}
\label{sec:transport}
The transport equation gives the rate of change of the distribution
function in time and in space due to the particle interactions. The distribution
function can in principle be solved from the transport equation. Here we limit
ourselves to cases valid for dilute systems where binary collisions dominate. 
In trying to study high energy nuclear reactions we need to know the properties
of many-particle system. While the properties of such a system depends upon the
interactions of the constituent particles and external constraints in kinetic
theory in the macroscopic level such a system is described by the net conserved
densities, the energy density, hydrodynamic velocity and dissipative quantities.
Thus the resulting fluid dynamic equations can be considered as an effective
kinetic theory. The results should be compared with other theories which goes
beyond dilute systems to check the deviations.

The relativistic Boltzmann transport equation for the invariant on-shell phase
space density $f(x,p)$ is 
\be
p^\mu\PD_\mu f(x,p)  = p^\m u_\m \D f(x,p) + p^\m \btd_\m f(x,p)
= {\mathcal{C}}[f(x,p)] \label{eq:RTE} ~,
\ee
where ${\mathcal{C}}[f]$ stands for the collision term. Here $\D= u^\m\PD_\m$ is
the covariant time derivative and $\btd_\m = \btu^{\m\n}\PD_\n$ is the covariant
gradient operator. Furthermore, $\btu^{\m\n} = g^{\m\n}-u^\m u^\n$ is the
projection tensor and $u^\m = (\gamma,\,\gamma \vec{v})$ is the 4-velocity 
where $\gamma =\sqrt{1-\tv^2}$
and the three--velocity of a particle is
$\tv \equiv \tp/p^0$. The four velocity is normalized such that $u^\m u_\m =1$.
The four momentum is $p^\m = (p^0, \tp)$ where $p^0 =E = 
\sqrt{\tp^2 +m^2}$ is the relativistic energy of the
particle
and $m$ is the mass of the particle.
Eq.~(\ref{eq:RTE}) describes the time
evolution of the single particle distribution function $f(x,p)$. 
The local equilibrium distribution function has the form
\be
f^\eqq (x,p) = A_0 {1\over e^{\beta_\mu p^\mu - \alpha}-a}
\label{eq:distrib}
\ee
where $A_0=g/(2\pi)^3$ and $a$ corresponds to the statistics of 
Boltzmann $(a = 0)$, Bose $(a = + 1)$, and
Fermi $(a = -1)$ distributions. Also the degeneracy  
$g=(2 J +1)$ where $J$ is the spin of the particle, 
$\beta_\mu = \beta u_\mu$, $\alpha = \beta \mu$ with $\beta=1/T$ the
inverse temperature, and $\mu$ is the
chemical potential. 

Let us define the first five moments of the distribution function $f(x,p)$ by
\bea
N^\m(x) &\equiv& \int f(x,p) p^\m \dd w~, \label{eq:N}\\
T^{\m\n}(x) &\equiv& \int f(x,p) p^\m p^\n \dd w~, \label{eq:T}\\
F^{\l\m\n}(x) &\equiv& \int f(x,p) p^\l p^\m p^\n \dd w~, \label{eq:F}\\
R^{\a\b\m\n}(x) &\equiv& \int f(x,p) p^\a p^\b p^\m p^\n \dd w~, \label{eq:R}\\
Q^{\a\b\m\n\r}(x) &\equiv& \int f(x,p) p^\a p^\b p^\m p^\n p^\r \dd w~, \label{eq:Q}
\eea
where 
\be
\dd w \equiv  {\dd^3 \tp\over p^0}.
\ee
For the equilibrium distribution function $f^{\eqq}(x,p)$,
Eq.~(\ref{eq:distrib}), the five moments can be written as
\bea
N^\mu_\eqq &=& \ci_{10}u^\m ~,\\
T^{\m\n}_\eqq &=& \ci_{20} u^\m u^\n - \ci_{21}\btu^{\m\n}~,\\
F^{\l\m\n}_\eqq &=& \ci_{30} u^\l u^\m u^\n - 3 \ci_{31} \btu^{(\l\m}u^{\n)}~,\\
R^{\a\b\m\n}_\eqq &=& \ci_{40} u^\a u^\b u^\m u^\n - 6 \ci_{41} \btu^{(\a\b}u^\m
u^{\n)} +3 \ci_{42} \btu^{(\a\b}\btu^{\m\n)}\\
Q^{\a\b\m\n\r}_\eqq &=& \ci_{50} u^\a u^\b u^\m u^\n u^\r - 10 \ci_{51}
\btu^{(\a\b} u^\m u^\n u^{\r)} +15 \ci_{52} \btu^{(\a\b}\btu^{\m\n} u^{\r)}
\eea
where the $\ci_{nk}$ are the equilibrium thermodynamic functions and are
presented in Appendix \ref{sec:momint}. Various contractions of the above 
equations produce useful relations such as 
\bea
u_\m N_{\eqq}^\m &=& \ci_{10} \equiv n~,~~~~~u_\m u_\n T_{\eqq}^{\m\n} = \ci_{20} \equiv \eps~,
~~~~~{T_{\eqq}}^\m_\m = \ci_{20} - 3\ci_{21} \equiv \eps-3p~, \label{eq:contra1}\\
u_\l {F_{\eqq}}^{\l\n}_\n &=& \ci_{30} - 3\ci_{31} = m^2 n~, 
~~~~~u_\a u_\b  {R_{\eqq}}^{\a\b\n}_\n = \ci_{40} - 3 \ci_{41}= m^2 \eps~. \label{eq:contra2}
\eea
Corresponding to the five moments are  the five auxiliary moments of 
$\Delta(x,p) f(x,p)$, which arises due to variations in the distribution
function,
\bea
\tilde{N}^\m(x) &\equiv& \int \Delta(x,p) f(x,p) p^\m \dd w ~, \label{eq:tN}\\
\tilde{T}^{\m\n}(x) &\equiv& \int \Delta(x,p) f(x,p) p^\m p^\n \dd w ~,\label{eq:tT}\\
\tilde{F}^{\l\m\n}(x) &\equiv& \int \Delta(x,p) f(x,p) p^\l p^\m p^\n \dd w
~,\label{eq:tF}\\
\tilde{R}^{\a\b\m\n}(x) &\equiv& \int \Delta(x,p) f(x,p) p^\a p^\b p^\m p^\n \dd w ~.
\label{eq:tR}\\
\tilde{Q}^{\a\b\m\n\r}(x) &\equiv& \int \Delta(x,p) f(x,p) p^\a p^\b p^\m p^\n
p^\r \dd w~, \label{eq:tQ}
\eea
where $\Delta(x,p) \equiv 1+a A_0^{-1}f(x,p)$.
For the equilibrium distribution function, Eq.~(\ref{eq:distrib}), and 
$\Delta^{\eqq}(x,p) f^{\eqq}(x,p)$ the auxiliary moments can be written as 
\bea
\tilde{N}^\mu_\eqq &=& \cj_{10}u^\m ~,\\
\tilde{T}^{\m\n}_\eqq &=& \cj_{20} u^\m u^\n - \cj_{21}\btu^{\m\n} ~,\\
\tilde{F}^{\l\m\n}_\eqq &=& \cj_{30} u^\l u^\m u^\n - 3 \cj_{31} \btu^{(\l\m}u^{\n)}
~,\\
\tilde{R}^{\a\b\m\n}_\eqq &=& \cj_{40} u^\a u^\b u^\m u^\n - 6 \cj_{41} \btu^{(\a\b}u^\m
u^{\n)} +3 \cj_{42} \btu^{(\a\b}\btu^{\m\n)}\\
\tilde{Q}^{\a\b\m\n\r}_\eqq &=& \cj_{50} u^\a u^\b u^\m u^\n u^\r - 10 \cj_{51}
\btu^{(\a\b}u^\m u^\n u^{\r)} +15 \cj_{52} \btu^{(\a\b}\btu^{\m\n} u^{\r)}
~.
\eea
where the  relativistic thermodynamic integrals $\cj_{nk}$  are also presented
in Appendix \ref{sec:momint}. Note that by Eq.~(\ref{eq:jnq}) the $\cj_{nk}$ can be related to the
$\ci_{nk}$, for example
\be
\cj_{21} = {\ci_{10}\over \b} \sks \cj_{31} = {(\ci_{20}+\ci_{21})\over \b} \sks 
\cj_{41} = {(\ci_{30}+2\ci_{31}) \over \b} \sks \cj_{42} = {\ci_{31}\over \b} ~.
\ee
Also by Eq.~(\ref{eq:dInq}) the $\cj_{nk}$ can be obtained as
differentiation of the $\ci_{nk}$ with respect to $\a$ and $\b$.

As in the case of the $\ci_{nk}$ similar contractions of the $\cj_{nk}$ leads
to similar relations, Eqs. (\ref{eq:contra1})--(\ref{eq:contra2}), with the
$\ci_{nk}$ now replaced by the $\cj_{nk}$.
For later reference we define the entropy 4-current in kinetic theory
as
\be 
S^\m(x) = -\int \dd w \,p^\m \left\{f(x,p)\ln\left(A_0^{-1} f(x,p)\right) - a^{-1}
A_0\Delta(x,p)\ln \Delta(x,p)\right\}~, \label{eq:S}
\ee
and we introduce a quantity 
\be
h_i \equiv {(\eps+p)\over n_i} = {w\over n_i} ={\cj_{31}\over \cj_{21}} ~,
\ee
where $w=\eps+p$, as the enthalpy per net conserved $i$th charge. Since in our
case we are considering one such charge, namely the net baryon number, $h$
denotes the enthalpy per net baryon. Similarly one defines $s/n$ as the entropy 
per net baryon.
%
\section{The 14-Field Theory of Non-equilibrium Relativistic Fluid Dynamics}
\label{sec:14-fields}
%
\subsection{Small deviations from thermal equilibrium}
\label{sec:deviations}
%
For a gas that departs slightly from the local thermal equilibrium, we may write
the distribution function as
\be
f(x,p) = f^\eqq (x,p)\left\{1+\Delta^\eqq(x,p)  \phi(x,p)\right\} 
\label{eq:noneqf}~,
\ee
where $\f(x,p)$ is the deviation function to be discussed in the next subsection. 
Substitution of Eq.~(\ref{eq:noneqf}) into 
Eqs.~(\ref{eq:N}), (\ref{eq:T}), and (\ref{eq:S}) yields
\bea
N^\mu(x) &=& N^\mu_\eqq (x) +\delta N^\mu(x) ~,\\
T^{\mu\nu}(x) &=& T^{\mu\nu}_\eqq (x) + \delta T^{\mu\nu}(x) ~,\\
S^\mu(x) &=& S^{\mu}_\eqq (x) + \delta S^\mu(x) ~,
\eea
where $N^\mu_\eqq $, $T^{\mu\nu}_\eqq$, and $S^\mu_\eqq $ have the ideal fluid (equilibrium) form,
and
\bea
\delta N^\mu(x) &=& \int f^\eqq(x,p)  \Delta^\eqq(x,p)  \phi(x,p) p^\mu \dd w ~,\\
\delta T^{\mu\nu}(x) &=& \int f^\eqq(x,p)  \Delta^\eqq(x,p)  \phi(x,p) p^\mu p^\nu \dd w ~.
\eea
For the entropy, expanding the term in the curly brackets in Eq.~(\ref{eq:S}) 
up to terms of second order in deviations $\f(x,p)$ (cf. Section \ref{sec:2nd}) leads to
\bea
\delta S^\mu(x) &=& S^\mu(x) - S^\mu_\eqq(x)  \nonumber\\
&&=
-\int\left[\alpha(x)-\beta_\nu(x) p^\nu\right]f^\eqq(x,p) \Delta^\eqq(x,p) \phi(x,p) p^\mu
\dd w \nonumber\\
&&-\half \int f^\eqq(x,p) \Delta^\eqq(x,p)  \left(\phi(x,p)\right)^2 p^\mu \dd w + \cdot\cdot\cdot 
\eea
which corresponds to the phenomenological expression
\be
\delta S^\mu(x) = -\alpha(x) \delta N^\mu(x) +\beta_\nu(x) \delta
T^{\mu\nu}(x) + Q^\mu(x) ~,
\label{eq:delS}
\ee
with
\be
Q^\mu(x) = -\half \int f^\eqq(x,p)\Delta^{\eqq}  (\phi(x,p))^2 p^\mu \dd w +\cdot\cdot\cdot ~.\label{eq:Qmu}
\ee
now explicitly defined in kinetic theory.
The five parameters $\alpha$ and $\beta_\mu$ which describe the nearby
equilibrium state can be specified by matching the equilibrium state to the
actual state. This is done by the prescription that the net charge density and
energy density are completely determined by the equilibrium distribution
function. That is, we impose the following conditions:
\bea
n &\equiv& \int \dd w p^\m u_\m f(x,p) 
= n_\eqq  \equiv \int \dd w p^\m u_\m f^\eqq (x,p)  ~,\\
\eps &\equiv& \int \dd w (p^\m u_\m)^2 f(x,p)  
= \eps_\eqq  \equiv  \int \dd w (p^\m u_\m)^2 f^\eqq(x,p)~.
\eea
Then $p(\eps,n) \equiv p(\eps_\eqq , n_\eqq )$ is the actual equation of
state. The above matching conditions also implies
\be
u_\mu \delta N^\mu = u_\mu u_\nu \delta T^{\mu\nu} = 0 ~. \label{eq:constraints}
\ee
It then follows from Eqs.~(\ref{eq:delS}) and (\ref{eq:Qmu}) that
\be
u_\mu \delta S^\mu = u_\mu Q^\mu \leq 0 ~,
\ee
confirming that equilibrium maximizes the entropy density under the above 
constraints, 
Eq.~(\ref{eq:constraints}), on particle and energy densities , and that $S_\eqq
^\mu$ and $T^{\mu\nu}_\eqq $ give the entropy and thermodynamical pressure
correctly to first order in deviations.
To fix the hydrodynamic velocity we may choose either the Eckart's or
Landau-Lifshitz' definition of the 4-velocity. The former case implies
\be
\btu^{\m\n} N_\n = \int \dd w \btu^{\m\n} p_\n f^\eqq(x,p) = 0~,
\ee
and the latter case implies
\be
\btu^{\m\n} T_{\n\s} u^\s = \int \dd w \btu^{\m\n} p_\n p_\s u^\s f^\eqq(x,p)
=0~.
\ee

In finding the equations for non-ideal fluid dynamics from microscopic/kinetic
theory there are basically two approaches which lead to linearized transport
equations: the Chapman--Enskog approximation \cite{Chapman} and   Grad's
14--moment approximation \cite{HG}. In the Chapman-Enskog method one  solves a
linearized Boltzmann equation for $f(x,p)$   which is linearized by assuming
that the function $\phi(x,p)$ that measures deviation from local equilibrium is
small. Terms which are non-linear in $\phi(x,p)$ and also the relative
variation over a mean free path are neglected.  One then uses the conservation
laws $\partial_\mu N^\mu \equiv \partial_\mu T^{\mu\nu} \equiv 0$ to express
the time derivatives $\D$ of the thermodynamical variables and the
four--velocity (hence $\D f^\eqq(x,p) $) in terms of spatial gradients
$\btd_\mu$ of $\alpha(x)$ and $\beta_\mu(x)$, correct to first order. 
Substitution into the kinetic expressions (cf. Eq.~(\ref{eq:RTE})) connects these
gradients to heat flow, diffusion and viscosity.  The second approximation
which is based on the moment method is more general and has a wide scope of
applicability. It will be discussed in the following section.

\subsection{Grad's 14--Moment Approximation}
\label{sec:grad}
In the phenomenological theories of relativistic non-ideal fluid dynamics by
Eckart \cite{CE} and by Landau and Lifshitz \cite{LL} (see \cite{AMI} for
details], 
instantaneous propagation of heat and viscous signals (acausality problem)
remained a puzzle  for many years.  However, in 1966, I. M\"uller \cite{IM}
traced the  origin of the difficulty to the neglect of terms of second order in
heat flux and viscosity in the conventional theory's expression for the
entropy. Restoring these terms, M\"uller was led to a modified system of
phenomenological equations which was consistent with the linearized form of
Grad's kinetic equations. M\"uller's theory was rediscovered and extended to
relativistic fluids by Israel and  Stewart \cite{IS} in the 1970's.

The progress made in phenomenology in getting rid of the acausality problem 
can also be made from kinetic theory.  The analogous paradox in
non-relativistic kinetic theory was resolved by Grad \cite{HG}, who showed in
1949 how transient effects could be treated by employing the method of moments
instead of the Chapman-Enskog normal solution. Suitable truncation of the
moment equations gave a closed system of differential equations which turned
out to be hyperbolic, with propagation speeds of the order of the speed of
sound. The relativistic version of Grad's method was developed by Israel and
Stewart \cite{IS}.

In phenomenology we seek a truncated hydrodynamical linearized description of
small departures from equilibrium in which only the 14 variables $N^\mu(x)$ and
$T^{\mu\nu}(x)$ appear. The microscopic counterpart of this is a truncated
description in which the function
\be
y(x,p) = \ln \left[A_0^{-1} f(x,p)/\Delta(x,p) \right] ~,
\ee
differs from any nearby local equilibrium value
\be
y_\eqq (x,p) = \alpha(x) -\beta_\mu(x) p^\mu ~,
\ee
by a function of momenta specifiable by 14 dynamic variables. 
The truncated description is accomplished by
postulating the relativistic Grad's 14--moment approximation
\cite{IS,HG}, or variational method \cite{deGroot}, that $y-y_\eqq $ can be
approximated by a quadratic function in momenta
\be
\f(x,p) \equiv y(x,p)-y_\eqq(x,p)  =\epsilon(x) -\epsilon_\mu(x) p^\mu +\epsilon_{\mu\nu}(x) p^\mu p^\nu + ...
~, \label{eq:phieq}
\ee
or
\be
y(x,p) = (\a+\epsilon(x)) -(\b_\m+\epsilon_\m (x)) p^\mu +\epsilon_{\mu\nu}(x) p^\mu p^\nu + ...
~,\label{eq:yeq}
\ee
where $\epsilon(x),\,\epsilon_\mu(x)$, and $\epsilon_{\mu\nu}(x)$ are small. 
Without loss of generality $\epsilon_{\mu\nu}(x)$ may be assumed traceless,
since its trace can be absorbed in redefinition of $\epsilon(x)$. The
non-equilibrium distribution function is given by Eq.~(\ref{eq:noneqf}) 
and it depends on the 14 variables $\alpha+\epsilon\,,\beta_\mu+\epsilon_\mu$,
and $\epsilon_{\mu\nu}$. While five of these determine the equilibrium state
the other nine variables are related to the dissipative fluxes. Inserting the
expression for $f(x,p)$, Eq.~(\ref{eq:noneqf}),  into the kinetic expressions for
$N^\mu$ and $T^{\mu\nu}$, Eqs.~(\ref{eq:N}) and (\ref{eq:T}), we then have
\bea
N^\mu &=& N^\mu_\eqq  +\epsilon \int \dd w f^\eqq(x,p)  \Delta^\eqq(x,p)  p^\mu \nonumber\\
&& - \epsilon_\nu \int \dd w f^\eqq(x,p)  \Delta^\eqq(x,p)  p^\mu p^\nu +
\epsilon_{\nu\l} \int \dd w f^\eqq(x,p)  \Delta^\eqq(x,p)  p^\mu p^\nu p^\l
 \label{eq:off-number}\\
T^{\mu\nu} &=& T^{\mu\nu}_\eqq  + \epsilon \int \dd w f^\eqq(x,p) 
\Delta^\eqq(x,p)  p^\mu p^\nu
-\epsilon_\l \int \dd w f^\eqq(x,p)  \Delta^\eqq(x,p)  p^\mu p^\nu p^\l
\nonumber\\
&&+
\epsilon_{\l \r} \int \dd w f^\eqq(x,p)  \Delta^\eqq(x,p)  p^\mu p^\nu
p^\l p^\r  \label{eq:off-energy-mom} \\
F^{\s\m\n} &=& F^{\s\m\n}_\eqq + \epsilon \int \dd w f^\eqq(x,p) 
\Delta^\eqq(x,p)  p^\s p^\m p^\n
-\epsilon_\l \int \dd w f^\eqq(x,p)  \Delta^\eqq(x,p)  p^\s p^\m p^\n p^\l
\nonumber\\
&&+
\epsilon_{\l \r} \int \dd w f^\eqq(x,p)  \Delta^\eqq(x,p) p^\s p^\m p^\n
p^\l p^\r  \label{eq:off-fluxes} 
\eea
Then by Eqs.~(\ref{eq:tN})-(\ref{eq:tR}) one can write the non-equilibrium part
of the number four current, the energy-momentum tensor and the fluxes as
\bea
\delta N^\m &=& \epsilon \tilde{N}^\mu_\eqq -\epsilon_\l
\tilde{T}^{\m\l}_\eqq
+\epsilon_{\l\n} \tilde{F}^{\l\m\n}_\eqq ~, \label{eq:del-N}\\
\delta T^{\m\n} &=& \epsilon \tilde{T}^{\m\n}_\eqq 
-\epsilon_\l \tilde{F}^{\l\m\n}_\eqq +\epsilon_{\l\r} \tilde{R}^{\l\r\m\n}_\eqq 
~, \label{eq:off-T}\\
\delta F^{\s\m\n} &=& \epsilon \tilde{F}^{\s\m\n}_\eqq -\epsilon_\l
\tilde{R}^{\s\m\n\l}_\eqq + \epsilon_{\l\r} \tilde{Q}^{\s\m\n\l\r}_\eqq ~.
\label{eq:off-F}
\eea
From the definitions of the dissipative fluxes, namely the particle drift flux
$V^\m = \btu^\m_\n \delta N^\n$, the energy flux $W^\m = \btu^\m_\n u_\r \delta
T^{\n\r}$, the heat flux  $q^\m = W^\m - h V^\m$, the bulk viscous  pressure 
$\Pi \equiv -{1\over 3} \btu_{\m\n} \delta T^{\m\n}$ and the shear stress 
tensor $\pi^{\m\n} \equiv \delta T^{\langle\m \n\rangle}$ and the matching
conditions, Eq.~(\ref{eq:constraints}), one then obtains the 14 variables 
$\alpha +\epsilon\,,\beta^\mu+\epsilon^\mu$, and $\epsilon^{\mu\nu}$  in terms
of the macroscopic fields   $n,\,\varepsilon,\,u^\mu,\,\Pi,\,q^\mu$, and
$\pi^{\mu\nu}$ \cite{IS}
\bea
\epsilon_{\mu\nu} &=& \ca_2(3u_\mu u_\nu-\btu_{\mu\nu})\Pi - \cb_1 u_{(\mu}q_{\nu)}
+\cc_0\pi_{\mu\nu} ~, \label{eq:tensor}\\
\epsilon_\mu &=& \ca_1 u_\mu \Pi -\cb_0 q_\mu ~,
\label{eq:vector}\\
\epsilon &=& \ca_0 \Pi ~. \label{eq:scalar}
\eea
The coefficients $\ca_i,\,\cb_i$ and $\cc_0$ are thermodynamic functions
given by
\bea
\ca_2 &=& {1\over 4 J_{42}\Omega} \,,~~~~~\cb_1 = {1\over \Lambda
\cj_{21}}\,,
~~~~~\cc_0 = {1\over 2 \cj_{42}} ~,\\
~~~~~\ca_1 &=& 3 \ca_2 D_{20}^{-1}\left\{ 4(\cj_{10}\cj_{41} - \cj_{20}
\cj_{31})\right\}\,,~~~~~\cb_0 =  \cb_1 {\cj_{41}\over \cj_{31}} ~,\\
~~~~~\ca_0 &=& 3 \ca_2 
D_{20}^{-1}\left\{ D_{30} + \cj_{41} \cj_{20} - \cj_{30} \cj_{31}\right\} 
~,
\eea
with
\bea
\Lambda &=& {D_{31}\over \cj_{21}^2} ~,\\
\Omega &=& -3\left({\partial \ln\ci_{31}\over \partial \ln\ci_{10}}\right)_{s/n}
+5 \nonumber\\
&=& -3{\cj_{31}\biggl(\cj_{21}\cj_{30} -\cj_{20}\cj_{31} \biggr) - 
        \cj_{41}\biggl(\cj_{21}\cj_{20}-\cj_{10} \cj_{31}\biggr)
	         \over \cj_{42} D_{20}} + 5
		 ~,\label{eq:om}\\
D_{nk} &=& \cj_{n+1,k} \cj_{n-1,k} - (\cj_{nk})^2  \label{eq:Dnq}~.
\eea

Once the deviation function is determined one can then derive the equations of 
motion for $n,\,\varepsilon,\,u^\mu,\,\Pi,\,q^\mu$, $\pi^{\mu\nu}$. 
In kinetic theory one
uses the moments of Boltzmann transport equation, Eq.~(\ref{eq:RTE}). 
For a general tensorial function of momenta $\Phi(p)$ we have 
\be
\int \dd w~ \Phi(p) p^\m \PD_\m f(x,p) = \int \dd w~ \Phi(p) \cc [f] ~.
\ee
For
Grad's 14--moments approximation we only need the equations for  the first
three moments of the distribution function $f(x,p)$. For $\Phi(p) =
1,\,p^\n$, and $p^\n p^\l$ we have respectively 
\bea
\int \dd w~ p^\mu \partial_\mu f(x,p) &\equiv& \partial_\mu N^\mu =
\int \dd w~ {\cal C} [f] \equiv 0 ~,\label{eq:Number}\\
\int \dd w~ p^\mu p^\nu \partial_\mu f(x,p) &\equiv& \partial_\mu
T^{\mu\nu} =
\int \dd w~ p^\nu{\cal C} [f] \equiv 0 ~,\label{eq:4-mom}\\
\int \dd w~ p^\mu p^\nu p^\l \partial_\mu f(x,p) &\equiv& \partial_\mu
F^{\mu\nu\l} =
\int \dd w~ p^\nu p^\l {\cal C} [f] = 
P^{\nu\l}~. \label{eq:Fluxes}
\eea
The first two moment equations give the 5 conservation laws, the particle
number and energy--momentum conservation. The remaining additional 9 equations
are obtained from the third moment equation which represents the balance of
fluxes. $F^{\mu\nu\l}$ is a completely symmetric tensor of fluxes and
$P^{\mu\nu}$ is its production density which is the rate of production per unit
4-volume of $\Phi(p)$ due to collisions. The balance equations are 15 in number
for only 14 fields.  The trace of $F^{\mu\nu\l}$ is mass squared times
Eq.~(\ref{eq:Number})  and $P^{\mu\nu}$ is traceless. That is,
\be
P^\nu_\nu = 0~~~~~\mbox{and}~~~~~F^{\mu\nu}_\nu = m^2 N^\mu ~,
\ee
so that the trace of the tensor equation, Eq.~(\ref{eq:Fluxes}), reduces to the
conservation law of particle numbers, Eq.~(\ref{eq:Number}). Among the 10
equations, Eqs.~(\ref{eq:Fluxes}), there are only 9 independent ones. 
The resulting relaxational transport equations \cite{IS} are presented in 
detail for applications in \cite{AMI}.
Thus the
set of equations, Eqs.~(\ref{eq:Number})--(\ref{eq:Fluxes}), is a set of 14
independent equations for 14 fields. In this 14-Field Theory (a field theory of
the 14 fields) for non-equilibrium relativistic fluid dynamics 
the dynamical equations, Eqs.~(\ref{eq:Number})-(\ref{eq:Fluxes}), with
representations (tensor decomposition) in the Eckart or particle frame
\bea
N^\m &=& n u^\m \sks\\
T^{\m\n} &=& \eps u^\m u^\n -(p(\eps, n)+\Pi) \btu^{\m\n} +2 q^{(\m} u^{\n)} +
\pi^{\lla\m\n\gra} \sks\\
P^{\m\n} &=& {4\over 3} \cc_\Pi\ca_2\left(3 u^\m u^\n-\btu^{\m\n}\right)\Pi 
+2\cc_q \cb_1 q^{(\m}u^{\n)} + {1\over 5}\cc_\pi \cc_0 \pi^{\lla\m\n\gra} \sks\\
F^{\m\n\l} &=& (\ci_{30} + [\ca_0 \cj_{30}-\ca_1 \cj_{40} +
3\ca_2(\cj_{50}+\cj_{51})]\Pi) u^\m u^\n u^\l \nonumber\\
&&-3(\ci_{31} + [\ca_0
\cj_{31}-\ca_1\cj_{41}+\ca_2(3\cj_{51}+5\cj_{52})]\Pi)\btu^{(\m\n}u^{\l)}
\nonumber\\
&&+3(\cb_0\cj_{41}-\cb_1\cj_{51}) q^{(\m}u^\n u^{\l)} 
-3(\cb_0\cj_{42}-\cb_1\cj_{52}) \btu^{(\m\n}q^{\l)} +6 \cj_{52} \cc_0
\pi^{(\m\n}u^{\l)} \sks
\eea
and the expressions for the $\ca_i$, $\cb_i$ and $\cc_i$ (known for a given
equation of state) gives a set of field equations for the variables
$n,\,\eps,\,u^\m,\,\Pi,\,\pi^{\lla\m\n\gra}$ and $q^\m$ which contains only three
positive valued functions of $n,\,\eps$, namely $\cc_\Pi,\,\cc_q,$ and
$\cc_\pi$.
The $\cc_A$'s are the collision integrals that depend on the microscopic
interactions such as cross-sections. The primary transport equations depends on 
the $\cc_A$'s and the local distribution function $f(x,p)$. The relaxation times
and length then follow from the product of the primary and the second order transport
coefficients. 
The resulting relaxational transport equations \cite{IS} are presented in in
detail for applications in \cite{AMI}.

In summary the 14-Field Theory of relativistic non-ideal fluids is concerned with the
conservation of net charge and of the energy-momentum and the balance of
fluxes. We rewrite the conservation and balance equations by splitting them into
spatial and temporal parts and by making right hand sides explicit. We have 
\bea
\PD_\m N^\m &=&0\sks
u_\n \PD_\l  T^{\n\l} =0\sks
\btu^\m_\n \PD_\l T^{\n\l} =0\sks \label{eq:conserv}\\
u_\n u_\l \PD_\g F^{\n\l\g} &=& -4\cc_\Pi \ca_2 \Pi \sks
\btu^\m_\n u_\l \PD_\g F^{\n\l\g} = \cc_q \cb_1 q^\mu\sks 
\PD_\l F^{\lla\m\n\gra\l} = {1\over 5} \cc_\pi \cc_0 \pi^{\lla\m\n\gra} ~.
\label{eq:balance}  
\eea

\section{$2^{\rm nd}$ order entropy 4-current in kinetic theory}
\label{sec:2nd}

The 14-Field theory is restricted by the requirement of hyperbolicity and
causality. These requirements can be deduced from the entropy principle. 
The kinetic expression for entropy, Eq.~(\ref{eq:S}), can be written as
\be
S^\mu = -\int \dd w~ p^\m~ \psi (f) ~, 
\label{eq:Ssec}
\ee
where
\bea
\psi(f) &=& \left[f(x,p)\ln\left(A_0^{-1}
f(x,p)\right) -a^{-1} A_0 \Delta(x,p) \ln \Delta(x,p) \right] \nonumber\\
&=&-a^{-1} A_0 \ln \Delta(x,p) +f(x,p)\ln \left[A_0^{-1} f(x,p)/\Delta\right] ~.
\eea
Expanding $\psi (f)$ around $\psi (f^\eqq)$  up to second order, i.e.,
\be
\psi(f) = \psi(f^\eqq) + \psi^{\prime}(f^\eqq)(f-f^\eqq) + \frac{1}{2}
\psi^{\prime\prime}(f^{\eqq})(f-f^\eqq)^2 + ... ~,
\ee
where 
\bea
\psi^{\prime}(f) &=& \ln \left[A_0^{-1} f(x,p)/\Delta(x,p)\right] ~, \\
\psi^{\prime\prime}(f) &=& \left[\Delta(x,p) A_0^{-1} f(x,p)\right]^{-1}~, 
\eea
are the first and second functional derivatives of $\psi(f)$ with respect to
$f(x,p)$, gives
\bea
\psi(f)&=& -a^{-1} A_0 \ln \Delta^\eqq(x,p) +\left[\alpha(x) - \beta_\mu(x)
p^\mu\right]
\left[f(x,p)-f^{\eqq}(x,p)\right] \nonumber\\
&&+\half \left[f^\eqq(x,p) A_0^{-1} \Delta^\eqq(x,p) \right]^{-1} 
(f(x,p)-f^\eqq(x,p) )^2 + ... ~.
\label{eq:psi}
\eea
Inserting the expression for $\psi(f)$, Eq.~(\ref{eq:psi}), into the expression 
for entropy flux, Eq.~(\ref{eq:Ssec}) yields
\be
S^\mu = S^\mu_\eqq  + \frac{q^\mu}{T} -\frac{1}{2} \int \dd w~p^\mu~
\psi^{\prime\prime}(f^\eqq )(f-f^\eqq )^2 ~,
\label{eq:S2nd}
\ee
where 
\be
S^\mu_\eqq  = p(\alpha,\beta) \beta^\mu - \alpha N_\eqq ^\mu +\beta_\l T_\eqq ^{\l\mu}
~,
\ee
is the equilibrium entropy and is obtained by inserting $f^\eqq (x,p)$,
Eq~(\ref{eq:distrib}), into Eq.~(\ref{eq:S}). 
Using Eqs.~(\ref{eq:tensor})--(\ref{eq:scalar}) for 
$\epsilon\,,\epsilon_\mu\,,$ and $\epsilon_{\l\mu}$ 
in terms of the fluxes, one substitute $f(x,p)$, Eq.~(\ref{eq:noneqf}), 
into Eq.~(\ref{eq:S2nd}) to get the non-equilibrium entropy 4--current,
\bea
S^\mu &=& S^\mu_\eqq  + \b q^\m  -\frac{1}{2}\b u^\m \left(\beta_0
\Pi^2 -\left[\beta_1 + w^{-1}\right]q^\l q_\l + \beta_2 \pi^{\nu\l}\pi_{\nu\l}
\right) \nonumber\\
&&-\b\left(\left[\alpha_0 + w^{-1}\right] q^\m\Pi 
-\left[\alpha_1 + w^{-1}\right]q_\l \pi^{\mu\l} \right)~.
\eea
The coefficients $\a_i,\,\,\b_i$ stand for [cf. \cite{IS}]
\bea
\alpha_0 &=& {D_{41} D_{20} - D_{31} D_{30}   \over \beta \Lambda  \Omega
\cj_{42} \cj_{21}\cj_{31} D_{20}} ~,\\
\alpha_1 &=& {\cj_{31} \cj_{52} - \cj_{41} \cj_{42}  \over \beta\Lambda \cj_{42} 
\cj_{21} \cj_{31}} ~,\\
\beta_0 &=&{3 \over \beta\cj_{42}^2\Omega^2} \left\{5 \cj_{52} - {3\over
D_{20}}[\cj_{31}(\cj_{31}\cj_{30}-\cj_{41}\cj_{20})+\cj_{41}(\cj_{41}\cj_{10}-\cj_{31}\cj_{20})]\right\}
~,\\
\beta_1 &=&{D_{41}\over \beta \Lambda^2 \cj_{21}^2 \cj_{31}} ~,\\
\beta_2 &=&\half {\cj_{52}\over \beta \cj_{42}^2} ~.
\eea
These seccond order coefficients involves the $\cj_{nk}$ which can be obtained
as differentiations of the $\ci_{nk}$ (or the equation of state $p\equiv
p(\a,\b)$) with respect to $\a$ and $\b$. The transition to $p\equiv p(\eps,n)$ is
effected by the relations
\be
n = \b{\partial p \over \partial \a}\sks \eps = -\left[p+\b{\partial p \over
\partial \b}\right] 
\ee
With the help of Gibbs' equation 
\be
\dd (p\b^\n) = N_{\eqq}^\n \dd \a -T_{\eqq}^{\m\n} \dd \b_\m ~,\label{eq:Gibbs}
\ee
the entropy production can be written immediately as in phenomenology. In
kinetic theory we can derive it  by substituting Eq.~(\ref{eq:noneqf}) into the
expression for entropy, Eq.~(\ref{eq:S2nd}). Then we invoke the  second law of
thermodynamics, $\PD_\m S^\m \geq 0$, and developing up to second order in the
deviation function. This leads to an expression for $T\PD_\mu S^\mu$ which can
be written as the sum of contribution $T\PD_\mu S^\mu_1$ linear  in a deviation
function $\phi(x,p)$ and another one $T\PD_\mu S^\mu_2$ which is quadratic in 
$\phi(x,p)$. 
Using Eqs.~(\ref{eq:tensor})--(\ref{eq:scalar}) 
for  $\epsilon,
\,\epsilon_\mu,\,$ and $\epsilon_{\mu\nu}$ in terms of dissipative
fluxes together with the help of the expression for the derivative of $f^\eqq(x,p)$
\be
p^\mu\PD_\mu (\ln A_0^{-1}\Delta^\eqq  f^\eqq ) = {p^\mu \PD_\mu f^\eqq  \over
f^\eqq \Delta^\eqq } = -p^\mu\PD_\mu\left\{{p^\nu u_\nu -\mu \over T}\right\} ~,
\ee
we obtain the following contributions 
\bea
T\partial_\mu S^\mu_1 &=& -\Pi\btd_\mu u^\mu
-q^\mu\left(\frac{\btd_\mu T}{T} -\D u_\mu\right) + \pi^{\mu\nu} 
\btd_{\langle\mu}u_{\nu\rangle} ~,\\
T\partial_\mu S^\mu_2 &=& -\frac{1}{2} \D\left\{\beta_0 \Pi^2
-\left[\beta_1 +w^{-1}\right]q^\mu q_\mu 
+\beta_2\pi^{\mu\nu}\pi_{\mu\nu} \right\} \nonumber\\
&&-\btd_\mu \left\{\left[\alpha_0 +w^{-1}\right]q^\mu \Pi 
-\left[\alpha_1 + w^{-1}\right] q_\nu \pi^{\mu\nu}\right\} ~.
\eea
The second law of thermodynamics requires the form of he collision term 
$\cc [f]$ to be such that the entropy production is given by a positive-definite
integral for any function $f(x,p)$. That is,
\be
\int \dd w~ p^\m \PD_\m \psi (f) \equiv -\PD_\m S^\m = \int \dd w~ \psi^\pr(f)
\label{eq:cfreq}
\leq 0 ~.
\ee
Eq.~(\ref{eq:cfreq}) together with the first two moment equations,
Eqs.~(\ref{eq:Number}) and (\ref{eq:4-mom}), and Eq.~(\ref{eq:yeq}) leads to
\be
0\leq \PD_\m S^\m = -\int \dd w~ \cc [f] y(x,p) = -\eps_{\m\n} P^{\m\n}
\ee
The entropy production can be written as 
\be
\b^{-1}\PD_\m S^\m = \z^{-1}\Pi^2 - \l^{-1} q^\a q_\a + (2\h)^{-1} \pi^{\a\b}
\pi_{\a\b} \geq 0 ~,
\ee
with
\be
\z = {\b\over 16 \ca_2^2\cc_\Pi} = {\beta \cj_{42}^2 \Omega^2 \over \cc_\Pi}\sks 
\l \equiv \k T = {\b \over \cb_1^2\cc_q } = 
{\beta \cj_{21}^2 \Lambda^2 \over \cc_q}\sks
\h = {5\b\over 2 \cc_0^2 \cc_\pi} = 10 {\beta \cj_{42}^2 \over \cc_\pi}
\ee
where the $\cc_A$ are the collision integrals which involves the cross-sections
of various processes. For the calculation of the primary transport
coefficients  see, for example, \cite{AMY}. For consistency the primary
transport coefficients, the second order transport coefficients (hence the
relaxation times and relaxation lengths) and the
equation of state have to be determined from the same model or theory.

From the kinetic theory approach the unknown phenomenological coefficients 
$\zeta,\, \k,\, \eta,\, \alpha_0,\,
\alpha_1,\, \beta_0,\, \beta_1,$ and $\beta_2$ can now be explicitly identified
from the knowledge of the collision term and the equation of state.  $\k$
stands for the thermal conductivity, and $\zeta$ and $\eta$ stands for the bulk
and shear viscous coefficients respectively. The transport coefficients involve
complicated collision integrals. The relaxation coefficients $\beta_0$,
$\beta_1$ and $\beta_2$ makes the theory a causal one. The coefficients
$\alpha_0$ and $\alpha_1$ arise from coupling between viscous stress and heat
flux.  Knowledge of the second order coefficients allows one to write the
primary transport  coefficients  in terms of the relaxation times.  Such
relaxation times depend on the collision term in the Boltzmann transport
equation, and their derivation is an extremely laborious task \cite{AMY,AM}. 
For present purposes, it suffices to know that the kinetic theory yields 
the form of the second-order entropy 4-current and of the evolution equations
of the  fluxes and that it provides the explicit values of the relaxation
coefficients in them.  
The relaxation times are related to the
transport coefficients multiplied by  the $\b_A$ and and the relaxation lengths
are related to the transport  coefficients multiplied by the $\a_A$, 
\bea
\t_\Pi &=& \zeta \b_0 \sks \t_q =\kappa T \b_1 \sks \t_\pi = 2\eta \b_2 \enspace, \label{eq:relt}\\
l_{\Pi q}&=& \zeta \a_0 \sks l_{q\Pi} =\kappa T\a_0 \sks
l_{q\pi} = \kappa T \a_1 \sks l_{\pi q} = 2\eta \a_1 ~. \label{eq:rell}
\eea
These are the relaxation times for the bulk pressure  ($\tau _\Pi $),  the heat flux
($\tau _q $), and the shear tensor ($\tau _\pi $) ; and the relaxation lengths for
the coupling between heat flux and bulk pressure ($l_{\Pi q}$, $l_{q\Pi}$) and
between heat flux and shear tensor ($l_{q\pi}$, $l_{\pi q}$). From these one can
already study the general dependence of the ratios of the transport coefficients
to the respective relaxation times or relaxation lengths,
\bea
{\zeta\over \t_\Pi} &=& {1\over  \b_0} \sks {\l_{qq}\over\t_q} ={1\over \b_1} 
\sks {\eta\over \t_\pi} = {1\over 2 \b_2} \enspace, \label{eq:coeft}\\
{\zeta\over l_{\Pi q}}&=& {\l_{qq}\over l_{q\Pi}} ={1\over \a_0} 
\sks 
{\l_{qq}\over l_{q\pi}} ={2\eta\over l_{\pi q}} = {1\over \a_1} \enspace, 
\label{eq:coefl}
\eea
where $\l_{qq}$ stands for $\kappa T$. 
The various time scales and length scales mention here are to be contrasted
with appropriate dynamical time scales and the transverse and longitudinal
length scales for a given nuclear collision. In high energy nuclear collisions
the time scales over which the various constituents of the matter produced in
these collisions equilibrate are important in understanding the nonequilibrium
effects. Thus a knowledge of relevant nonequilibrium properties is essential
for a complete description. In the collision dominated regime the mean
collision times and the relaxation times associated with the changes in
distribution functions provides the information about the trends towards global
dynamics. These times are generally smaller than the times associated with the
transport of momentum and energy.

Once the entropy 4-vector is obtained from the kinetic theory one may then
construct a generating function which may be a 4-vector \cite{LMR} or a scalar
\cite{GL}. Thus from the entropy 4-vector one constructs/define a 4-vector
$\Psi\equiv\Psi(\epsilon^\pr, \epsilon_\l^\pr,\epsilon_{\n\l})$
\be
\Psi^\m \equiv S^\m +\epsilon^\pr N^\m -\epsilon_\l^\pr T^{\m\l}
+\epsilon_{\n\l} F^{\m\n\l} \label{eq:Psim}
\ee
where $\epsilon^\pr=\a+\epsilon$ and $\epsilon_\l^\pr = \b_\l +\epsilon_\l$.
From the differential form of $\Psi^\m$
\be
\dd \Psi^\m = N^\m \dd \epsilon^\pr - T^{\m\l} \dd \epsilon_\l^\pr + F^{\m\n\l}
\dd \epsilon_{\n\l} \label{eq:dPsim}
\ee
and the entropy principle (the second law of thermodynamics) which is written as
\be
\PD_\m S^\m +\epsilon^\pr\PD_\m N^\m -\epsilon_\l^\pr\PD_\m T^{\m\l} 
+\epsilon_{\n\l}(\PD_\m F^{\m\n\l} - P^{\n\l}) \geq 0 \label{eq:sec-law}
\ee
one then obtains
the functions $N^\m$, $T^{\m\n}$, and $F^{\lla\n\l\gra \m}$ as functions of
$\epsilon^\pr,\,\epsilon_\l^\pr,\,\epsilon_{\n\l}$. That is,
\be
N^\m = {\PD \Psi^\m \over \PD \epsilon^\pr}\sks
T^{\m\n} = -{\PD \Psi^\m \over \PD \epsilon_\l^\pr}\sks
F^{\lla\n\l\gra \m} = {\PD \Psi^\m \over \PD \epsilon_{\n\l}}-{1\over 4}
g^{\n\l} g_{\a\b} {\PD \Psi^\m \over \PD \epsilon_{\a\b}}\sks
-\epsilon_{\m\n} P^{\m\n} \geq 0\label{eq:restrict}
\ee
Using the definition of $\Psi^\m$, Eq.~(\ref{eq:Psim}), we can convert
Eq.~(\ref{eq:dPsim}) into a differential form for the entropy 4-vector $S^\m$
\be
\dd S^\m = -\epsilon^\pr \dd N^\m + \epsilon_\l^\pr \dd T^{\n\l}
-\epsilon_{\n\l} \dd F^{\m\n\l}
\ee

In discussing the constraints imposed on the dynamical equations by the entropy
principle, hyperbolicity and causality requirements we need to cast the system
of dynamical equations in a more transparent form. We introduce  
$Y^\m_A = \{N^\m, T^\m_\n, F^\m_{\lla\n\l\gra}\}$ representing the primary dynamical
variables, $P^\m_A = \{0,0,P_{\n\l}\}$ representing dissipation source tensor
and 
$X_A = \{\epsilon^\pr, \epsilon_\l^\pr,\epsilon_{\n\l}\}$ representing the
auxiliary dynamical variables. 
$A = 1, \cdots ,14$. Then the dynamical equations
Eqs.~(\ref{eq:Number})-(\ref{eq:Fluxes}) with the restrictions
Eqs.~(\ref{eq:restrict}) can be written in the form
\be
\PD_\m Y^\m_A = P_A\sks \mbox{and}~~~~ Y^\m_A = {\PD \Psi^\m \over \PD X_A}
\ee
which can be combined to form a symmetric form
\be
{\PD^2 \Psi^\m \over \PD X_A \PD X_B} u_\m \dot{X}_B 
- {\PD^2 \Psi^\m \over \PD X_A \PD X_B} \btd_\m X_B = P_A \label{eq:symsys}
\ee
For the system of equations Eqs.~(\ref{eq:symsys}) to be hyperbolic (i.e., has
well-posed initial value formulation) and causal (hyperbolic with no fluid
signals propagating faster than light)  we require
that
\be
 {\PD^2 \Psi^\m \over \PD X_A \PD X_B} u_\m 
\ee
be negative-definite for the physical states of the fluid.

The requirements of non-negative entropy production and of hyperbolicity imply
the following restrictions on the coefficients $\cc_\Pi,\, \cc_q,\, \cc_\pi$
and $\b_0,\,\b_1,\,\b_2$ respectively. These read
\bea
&& \cc_\Pi \geq 0\sks \cc_q \geq 0\sks \cc_\pi \geq 0 \label{eq:cond1}\\
&& \b_0 \geq 0\sks \b_1 \geq 0\sks \b_2 \geq 0 \label{eq:cond2}\\
&& {\PD p\over \PD n} >0\sks {\PD {\eps\over n} \over \PD T} >0 ~.\label{eq:cond3}
\eea
The conditions Eq.~(\ref{eq:cond1}) ensure the positivity of bulk viscosity,
shear viscosity and heat conductivity. The conditions Eq.~(\ref{eq:cond2})
ensure that the entropy density has its maximum in equilibrium and together with
conditions Eq.~(\ref{eq:cond1}) they also 
ensure the positive relaxation times for $\Pi,\,q^\m$, and $\pi^{\lla\m\n\gra}$. 
The two
inequalities in Eq.~(\ref{eq:cond3}) are the stability conditions on
compressibility and specific heat. The whole set of inequalities
Eqs.~(\ref{eq:cond1})-(\ref{eq:cond3}) ensures finite speeds. Some of the
consequences of the requirement of hyperbolicity impose the bounds on the values
of $\Pi,\,q^\m$, and $\pi^{\lla\m\n \gra}$ in terms of the independent variables
of the equation of state e.g., $p$ and $\eps$. The hyperbolicity requirement,
that is the requirement that the field equations form a symmetric hyperbolic
system ensures that the problem is well posed and that the characteristic speeds
are real and finite. This requirement is equivalent to the requirement that the
second differential $\delta^2 S^\m(n,\eps,\Pi,q^\m,\pi^{\m\n})$ of the entropy must be negative definite.

\section{The equation of state prescription}
\label{sec:EoS}
%
For proper description of the space-time evolution of the hot and dense 
nuclear matter, produced in high energy nuclear collisions, using fluid
dynamics, one needs an equation of state to close the system of evolution
equations. The pressure as a function of energy density and baryon density,
$p(\eps,n)$, is required for solving the fluid dynamical equations numerically.
It is convenient to tabulate this function, since calculating the pressure in
the hadron phase for given $\eps$ and $n$  requires a double root search to
find $T$ and $\mu$.  To facilitate fast and easy numerical computation one
discretizes the  $\eps-n$ plane.  Every other thermodynamic quantity can be
calculated as a function of $\eps$ and $n$. Of particular importance are
temperature, baryochemical potential, entropy density, the relaxation and
coupling coefficients. For the
fluid dynamical calculations intermediate values of thermodynamic quantities are
calculated by two--dimensional linear interpolation.

For this case study the nuclear matter is described by a $\s-\o$-type model
\cite{walecka} for the hadronic matter phase and by the MIT bag model
\cite{MIT} for the quark-gluon plasma, with a first-order phase transition
constructed via the Gibbs phase equilibrium conditions. This type of an
equation of state was presented in \cite{DHR,migdhr}. 

In the hadronic matter, the equation of state, i.e., the pressure $p$ as a
function of the independent thermodynamical
variables temperature $T$ and baryochemical potential $\mu$, 
for non-strange
interacting nucleonic matter is defined by \cite{gori}
\begin{eqnarray} \label{phad}
p_{had}(T,\mu) & = & p_N(T,\nu;M^*) + p_N(T,-\nu;M^*) + \sum_i p_i(T;m_i)\\
         & + & n\, {\cal V}(n) - \int_0^n {\cal V}(n')\, {\rm d}n'
               - \rho_s\, {\cal S}(\rho_s) + \int_0^{\rho_s}
               {\cal S}(\rho_s')\, {\rm d}\rho_s'~. \nonumber
\end{eqnarray}
Here,
\begin{equation} \label{pn}
p_N(T,\nu;M^*) = \ci^N_{21}(\f^*,z^*)~, ~~~~~ 
p_N(T,-\nu;M^*) = \ci^N_{21}(-\f^*,z^*)~,
\end{equation}
where $\f^*\equiv\b\n$ and $z^*\equiv\b M^*$, is the pressure of an ideal gas
of nucleons moving in the scalar potential
${\cal S}$ and the vector potential ${\cal V}$. These potentials generate an
{\em effective\/} nucleon mass
\begin{equation}
M~~ \longrightarrow~~ M^* \equiv M - {\cal S}(\rho_s)~,
\end{equation}
where $M= 938$ MeV is the nucleon mass in the vacuum, and also shift the
one--particle energy levels
\begin{equation}
E^{\pm} \equiv \sqrt{{\bf p}^2 + M^2}~~\longrightarrow~~ E^* \pm {\cal V}(n) \equiv
\sqrt{{\bf p}^2 + \left(M^*\right)^2} \pm {\cal V}(n)~.
\end{equation}
The vector potential is conveniently absorbed in the {\em effective\/}
chemical potential
\begin{equation}
\nu \equiv \mu - {\cal V}(n)~,
\end{equation}
giving rise to the interpretation of Eq.~(\ref{pn})
as the pressure of an ideal gas of  quasi-particles with mass
$M^*$ and chemical potential $\nu$.
Furthermore,
\begin{equation}
p_i(T;m_i) = \ci^i_{21}(0,z_i)~,
\end{equation}
where $z_i\equiv \b m_i$, is the pressure of an ideal gas of mesons with degeneracy $g_i$
and mass $m_i$. Only the pions 
($g_{\pi}=3,\, m_{\pi}=138$ MeV) are considered since
they are the lightest and thus most abundant mesons. 
$n$ is the net baryon density,
\be \label{n}
n(T,\mu)  \equiv  \left. \frac{\partial p}{\partial \mu} \right|_T \\
   = \ci^N_{10}(\f^*,z^*)-\ci^N_{10}(-\f^*,z^*)~,
\ee 
and
\begin{equation}
\rho_s(T,\mu) \equiv \frac{g_N}{(2\pi)^3} \int {\rm d}^3
{\bf p}~\frac{M^*}{E^*}~ \left[ \frac{1}{e^{(E^* - \nu )/T}+1}
              + \frac{1}{e^{(E^* + \nu )/T}+1} \right]
\end{equation}
is the scalar density of nucleons.

Once the pressure is known, the entropy and energy density can be obtained
from the thermodynamical relations,
\begin{equation} 
s = \left. \frac{\partial p}{\partial T} \right|_{\mu}~,~~\eps =
Ts + \mu n - p~.\label{eq:thdyn}
\end{equation}
The potentials ${\cal V,S}$ are specified as in Ref.\ \cite{migdhr} 
\begin{equation} \label{choice}
{\cal V}(n) = C_V^2\, n - C_d^2\, n^{1/3}~,~~{\cal S}(\rho_s) = C_S^2\,
\rho_s~,
\end{equation}
where  $C_V^2= 238.08\, {\rm GeV}^{-2},\,
C_S^2 = 296.05\, {\rm GeV}^{-2},\, C_d^2 = 0.183$, are the parameters which 
which leads to the reasonable values for the effective mass and the
compressibility of $M^*_0 = 0.635\, M,\, K_0 = 300\, {\rm MeV}$.
$T$ and $\mu$ emerge naturally from the
double root search. 
Then 
using the results from Appendix \ref{sec:momint}, i.e., Eqs.~(\ref{eq:inq}) and
(\ref{eq:jnq}) or  Eqs.~(\ref{eq:inq2}) and (\ref{eq:jnq2}), one calculates the
thermodynamic integrals $\ci_{nk}$ and $\cj_{nk}$ for given $\eps$ and $n$
Then the $\a_i(\eps,n)$ and $\b_i(\eps,n)$ are determined. 

In the quark gluon plasma phase we employ the standard MIT bag model \cite{MIT}
equation of state 
for massless, non-interacting quarks and gluons, i.e.,
\begin{equation} \label{pQGP}
p_{QGP}(T,\mu) = \ci_{21}^{\pm}(0,\f) -B~,
\end{equation}
where  $\ci^{\pm}_{21}$ is given in Appendix \ref{sec:ultra-rel} and $B$ is the
bag constant. Other thermodynamical quantities follow again from Eqs.~(\ref{n}) and 
(\ref{eq:thdyn}).
Note that $p$ does not depend explicitly on $n$ for this EoS,
\begin{equation} \label{pQGP2}
p = \frac{1}{3} (\ci^{\pm}_{20} - 4B)~.
\end{equation}
The value of the bag constant is taken to be $B=(235\, {\rm MeV})^{4}$ 
which results in a phase transition temperature of $T_c \simeq 169$ MeV at
vanishing baryon density. 
Other thermodynamical quantities again follow from Eq.~(\ref{eq:thdyn}).
The baryon chemical potential $\mu_B$ is related to $\mu_q$ by
$\mu_B=3\mu_q$ and the net baryon charge density $n_B$ is related to 
the quark--gluon plasma baryon charge density $n_b$ by $n_B = n_b/3$. 
In the quark--gluon plasma the $n=0$ case is simple since we also have $\mu=0$,
and one obtains from Eqs.~(\ref{pQGP}) and (\ref{pQGP2}) the simple formula 
for the temperature, $T= [60 \eps/(32+21 N_f) \pi^2]^{1/4}$. $N_f$ is the number
of quark flavours. In our equation of state we consider $N_f=2$. 
For finite $\mu$, $T$ can be eliminated from the equation of $\eps$ using
the equation for $n$. This results in a sixth order equation in
$\mu$,
\begin{equation} \label{sixt}
0 = \frac{6N_f-8}{1215 \pi^2}~\mu^6 - \frac{21N_f-58}{30}~n \mu^3 + \eps \mu^2
- 27\frac{(32+21N_f) \pi^2}{20N_f^2}~n^2~,
\end{equation}
which has to be solved numerically. After that, $T = [9n/N_f \mu -
\mu^2/9\pi^2]^{1/2}$ (which follows from the equation for $n$).  Once $T$ and
$\m$ are known for given $\eps$ and $n$, the thermodynamic integrals $\ci_{nk}$
and $\cj_{nk}$ are also calculated as functions of $\eps$ and $n$ using the
results of Appendix~\ref{sec:ultra-rel} and thus the second-order coefficients
$\a_i(\eps,n)$ and $\b_i(\eps,n)$ are also calculated. 

In the mixed phase the quark gluon plasma phase equation of state 
Eq.~(\ref{pQGP}) is matched to the hadronic phase equation of state 
Eq.~(\ref{phad}) using the Gibbs' phase equilibrium conditions,
\be \label{gibbs}
p_{had} = p_{QGP}, ~~~~~T_{had} = T_{QGP}, ~~~~~\m_{had} = \m_{QGP}~.
\ee
In the mixed phase, for given temperature $T$ and chemical potential $\m$ 
the values of energy and baryon density read
\bea 
\eps &=& \l_Q\eps_Q(T,\m) + (1-\l_Q)\eps_H(T,\m)~,\\
n &=& \l_Q n_Q(T,\m) + (1-\l_Q) n_H(T,\m)~,
\eea
where $\l_Q$ is the fraction of volume the quark gluon plasma occupies in 
the mixed phase. Conversely, for given $\ep,~n$ these equations yield values 
for $\l_Q,~ T,~\m$.
This is done numerically using the values of
$\eps_{H,Q}(T,\m),~n_{H,Q}(T,\m)$. Once $T$ and $\m$ are known the pressure
follows from Gibb's phase equilibrium condition,
Eq.~(\ref{gibbs}). Similarly the other thermodynamic quantities can be calculated
as function of $\eps$ and $n$. Of particular importance are temperature,
baryo-chemical potential, entropy density and the rest of the thermodynamic
integrals, $\ci_{nk}$ and $\cj_{nk}$, from which the second-order coefficients
$\a_i(\eps,n)$ and $\b_i(\eps,n)$ are evaluated.

\begin{figure}[hp]
\centering
\epsfig{figure=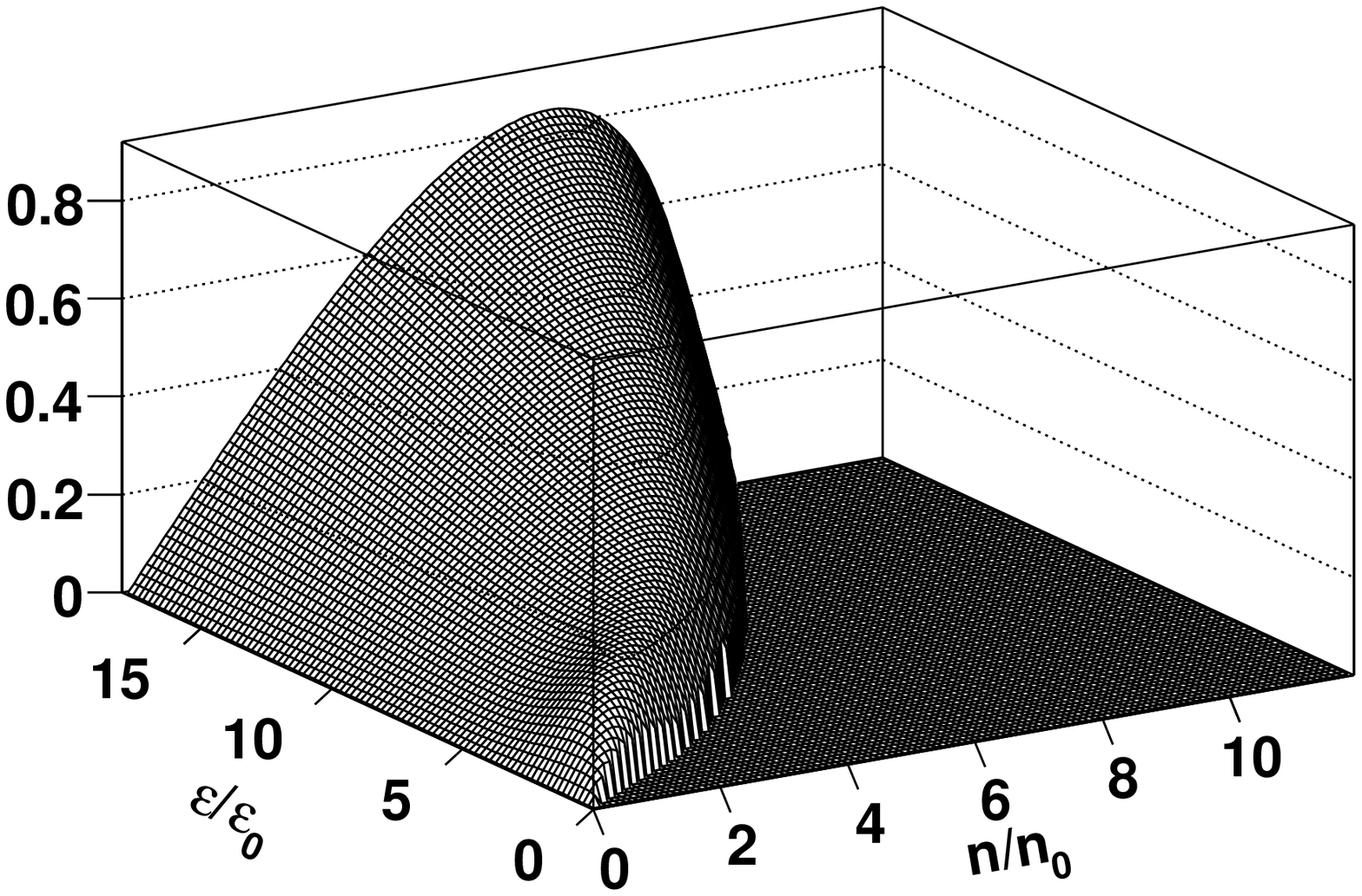, width=16.0 cm}
\caption{ The energy density and net baryon density dependence of $\ci_{30}$ 
(in units of GeV$^{2}$.fm$^{-3}$) for the hadronic part of the equation of state. The energy density and net
baryon densities are in units of the ground state densities. $n_0$ = 0.16
fm$^{-3}$ and $\eps_0$ = 0.147 GeV fm$^{-3}$.}
\label{fig:i30}
\end{figure}

\begin{figure}[hp]
\centering
\epsfig{figure=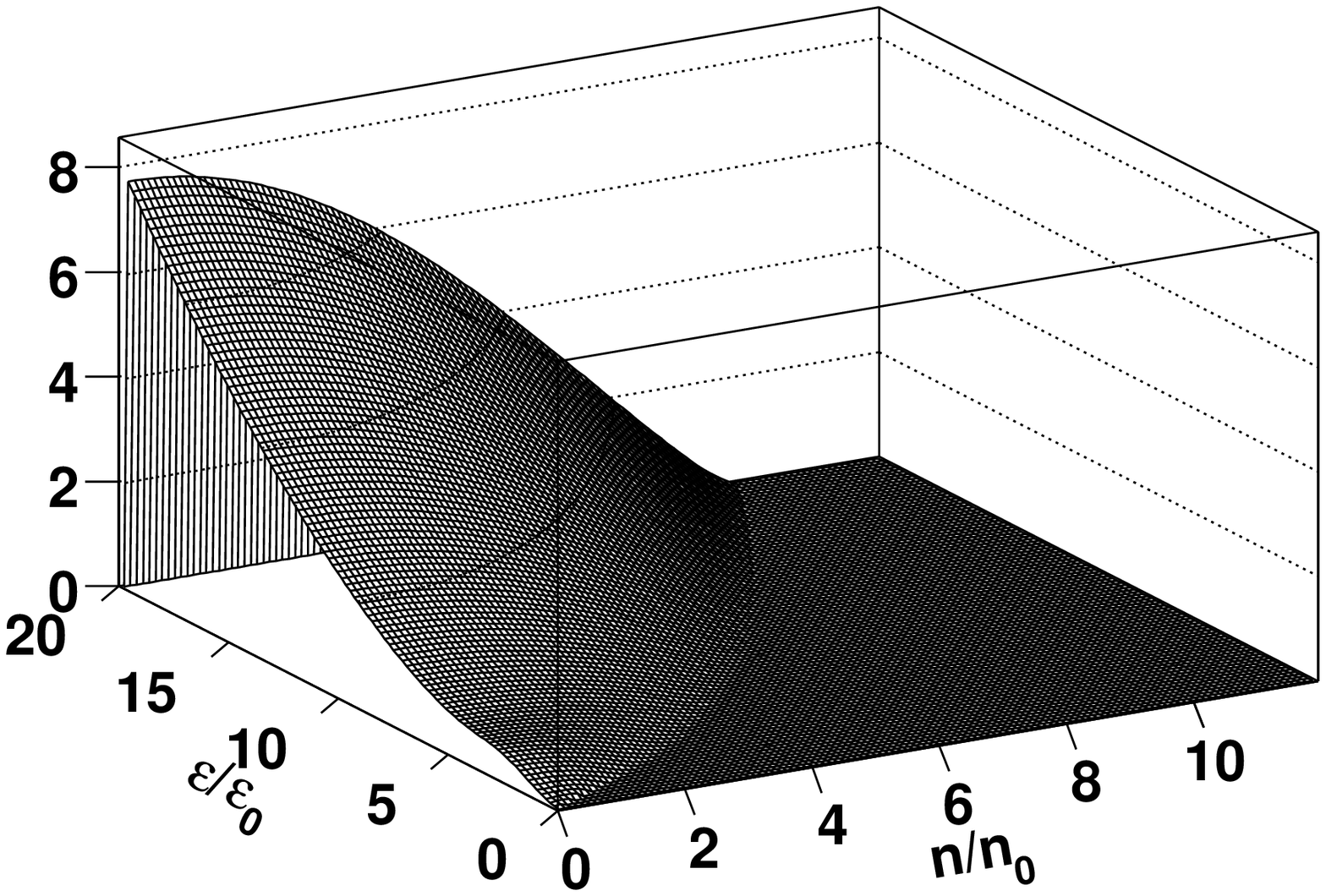, width=16.0 cm}
\caption{ The energy density and net baryon density dependence of $\cj_{50}$ 
(in units of GeV$^{4}$.fm$^{-3}$) for the hadronic part of the equation of state. The energy density and net
baryon densities are in units of the ground state densities. $n_0$ = 0.16
fm$^{-3}$ and $\eps_0$ = 0.147 GeV fm$^{-3}$.}
\label{fig:j50}
\end{figure}

In Figs.~\ref{fig:i30} and \ref{fig:j50} we show the dependence of the
relativistic thermodynamic integrals on this particular equation of state. The
results are only for the hadronic phase without the phase transition into the
quark gluon plasma. The results of the latter scenario will be explored in full
detail somewhere else.

\section{Freeze-Out and Particle Spectra}
\label{sec:freeze}
%
%
At any space--time point of the
many--particle (fluid dynamics) evolution a particle ``freezes out'' when its interactions with the rest of the system
cease. This depends on the  mean free path of the
particular particle. If the mean free path is small enough such that local 
thermodynamical equilibrium is established, fluid dynamics is valid. For larger
mean free path non-equilibrium effects become increasingly important, and if
the  mean free path exceeds the system's dimension, the latter starts to
decouple into free-streaming particles. To describe the system's evolution in
the latter two stages in principle requires kinetic theory.

The region where the mean free path is of the order of the system size can be approximated 
by a hypersurface in  space-time \cite{Landau,CF}.  When a fluid
element crosses this hypersurface, particles contained in that element freeze
out instantaneously. The freeze-out should be treated in a
self-consistent way taking into account the non-equilibrium effects caused by
particles leaving the system. 

We will consider isothermal freeze-out which corresponds to freeze-out at a
constant  fluid temperature $T_f$. One could also consider freeze-out at a
constant center of mass  time $t_f$. In the later scenario the temperature
along the hypersurface is no longer constant. 

In the calculations of particle spectra at freeze-out, the distribution of
particles that have decoupled from the fluid is determined as follows. The
Lorenz-invariant momentum-space distribution of particles crossing a
hypersurface $\Sigma$ in Minkowski space is given by \cite{CF}
\be
E{\dd N\over \dd^3 {\tp}} = \int_\Sigma \dd \Sigma_\m p^\m f(x,p)~,
\label{eq:cf}
\ee
where $\dd \Sigma_\m$ is the normal vector on an infinitesimal element of the
hypersurface $\Sigma$. $\dd \Sigma_\m$ is naturally chosen to point outwards with
respect to the hotter interior of $\Sigma$ since Eq.~(\ref{eq:cf}) is supposed
to give the momentum distribution of particles decoupling from the fluid. 
The distribution function $f(x,p)$ is given
by Eq.~(\ref{eq:noneqf}).

The calculations simplifies in the case of longitudinal boost invariance and
transverse translation invariance \cite{Bj}. In this case the on-shell phase
space distribution function depends only on the reduced phase space variables
$(\t,\c,\tp_\perp)$, where $\t^2 = t^2 - z^2$ and $\c = \h-y$, in terms of the
(longitudinal) particle rapidity $y=\tanh^{-1} (p^z/p^0)$ and the (longitudinal)
fluid rapidity $\h = \tanh^{-1}(z/t)$.
The collective flow velocity field, the momentum and the heat flux 
in this case are given by 
\bea
u^\m &=& x^\m/\t = (\cosh \h,{\ten 0}_\perp, \sinh \h)~, \\
p^\m &=& (m_\perp \cosh y,p_\perp\cos\f_p,p_\perp\sin\f_p,m_\perp\sinh y)~,\\
q^\m &=& q l^\m = q(\sinh \h, {\ten 0}_\perp, \cosh \h)~,
\eea
where $l^\m$ is a space-like 4-vector $l^\m l_\m=-1$ and it is orthogonal to the
4-velocity $l^\m u_\m=0$, $\phi_p$ is the angle of particle's momentum around
the $z$-axis and $m_\perp = \sqrt{\vec{p}_\perp^2+m^2}$ is the transverse mass which
reduces to $p_\perp$ for the massless particles. The shear tensor is given by
\be
\pi^{\m\n} = 
    \begin{pmatrix}  
    \pi_s\sinh^2 \h        &\quad 0   &\quad 0
                      &\quad \pi_s\cosh\h\sinh\h 
\\
    0  &\quad -{\pi_s\over 2}       &\quad 0  
                      &\quad 0  
\\
    0 &\quad 0    &\quad -{\pi_s\over 2}        
                      &\quad 0  
\\ 
    \pi_s\cosh\h\sinh\h &\quad 0   &\quad 0 
                      &\quad \pi_s\cosh^2\h 
    \end{pmatrix}  ~,
\ee
where $\pi_s$ is the scalar shear pressure (to distinguish it from the constant
$\pi$).
In our particular case at freeze-out $\t = \t_f$, $\Sigma$ is conveniently
parameterized by $\Sigma^\m = (\t_f\cosh \h,\tx_\perp,\t_f\sinh\h)$ with
$\dd \Sigma^\m = (\t_f\cosh \h,{\ten 0}_\perp,-\t_f\sinh\h) \dd \h \dd {\ten
x}_\perp$. Therefore, $\dd \Sigma^\m p_\m = \t_f m_\perp \cosh\c \dd\h \dd
\tx_\perp$  and  $p^\m u_\m = m_\perp\cosh \c$. 

The distribution function however now includes the corrections due to
dissipative effects. Now there is the equilibrium part and non-equilibrium part.
With the expressions, Eqs.~(\ref{eq:tensor})--(\ref{eq:scalar}), 
for $\epsilon(x)$,
$\epsilon_\n(x)$ and $\epsilon_{\m\n}(x)$ in terms of the dissipative fluxes we
can write the deviation function as
\be
\f(x,p) = \left[\ca_0 - \ca_1 (p^\m u_\m) + \left[4 (p^\m u_\m)^2-m^2\right]\ca_2\right]\Pi 
+ \left[\cb_0-\cb_1(p^\m u_\m)\right](p^\m q_\m) +\cc_0 p^\m p^\n\pi_{\m\n} + \cdot\cdot\cdot ~.
\ee 
In our present freeze-out prescription we have  
\bea
p^\m q_\m &=& q m_\perp \sinh \c ~,\\
p^\m p^\n \pi_{\m\n} &=&  \pi_s m_\perp^2\sinh^2 \c - {1\over 2}\pi_s
p_\perp^2~,
\eea
and the deviation function becomes
\bea 
\f(x,p) &=& \left[\ca_0 - \ca_1 (m_\perp\cosh\c) +
[4 (m_\perp\cosh\c)^2-m^2]\ca_2\right]\Pi \nonumber\\
&&+ \left[\cb_0-\cb_1 (m_\perp\cosh\c)\right](m_\perp\sinh\c)q \nonumber\\
&&+\cc_0 \left[ m^2_\perp\sinh^2\c - {1\over 2}p_\perp^2\right]\pi_s ~.
\label{eq:1dphi}
\eea
In the freeze-out prescription of a system with longitudinal boost invariance
and transverse cylindrical symmetry \cite{MR} one frequently uses the 
integrals of the form
\bea
\chi_{ijmn}^{(1)} &=& \int_0^{2\pi} \dd\psi \sin^i\psi \cos^j\psi \int_0^\infty\dd
\chi \sinh^m\chi \cosh^n \chi \nonumber\\
&&\times {1\over e^{z_\perp\cosh\chi - \alpha_\perp\cos\psi
-\f}-a}~,\label{eq:chi1}\\
\chi_{ijmn}^{(2)} &=& \int_0^{2\pi} \dd\psi \sin^i\psi \cos^j\psi \int_0^\infty\dd
\chi \sinh^m\chi \cosh^n \chi \nonumber\\
&&\times {e^{z_\perp\cosh\chi - \alpha_\perp\cos\psi-\f}\over \left[e^{z_\perp\cosh\chi - \alpha_\perp\cos\psi
-\f}-a\right]^2}~,\label{eq:chi2}
\eea
where $\psi = \phi-\phi_p$ is given in terms of the azimuthal angle of the
surface element of the fluid $\phi$ and the angle of the particle's momentum
$\phi_p$  around $z$, $z_\perp = \g_\perp m_\perp/T$ and $\a_\perp=\g_\perp
v_\perp p_\perp/T$. 
The powers in the trigonometric and hyperbolic functions
comes from the product of the volume element parameterization in cylindrical
coordinates with boost invariance and the distribution function $f(x,p)\equiv
f(p_\m u^\m,\Pi,p_\m q^\m, p_\m p_\n \pi^{\m\n})$ as given by Eq.~(\ref{eq:noneqf}).
Eq.~(\ref{eq:chi1}) arises due to the first term of Eq.~(\ref{eq:noneqf}) and this
gives the equilibrium (ideal fluid) contribution to the spectra while
Eq.~(\ref{eq:chi2}) arises from the second term of Eq.~(\ref{eq:noneqf}) and it
gives non-equilibrium (non-ideal fluid) contribution to the spectra. Expanding
the Fermi-Dirac and Bose-Einstein distribution functions in geometric series we
can write Eqs.~(\ref{eq:chi1}) and (\ref{eq:chi2}) as
\bea
\chi_{ijmn}^{(1)} &=& \sum_{k=1}^\infty \left(\mp\right)^{k-1} e^{k\f}
\nonumber\\
&&\times \int_0^{2\pi} \dd\psi \sin^i\psi \cos^j\psi  e^{k\a_\perp \cos\psi}\int_0^\infty\dd
\chi \sinh^m\chi \cosh^n \chi  e^{-k z_\perp\cosh\chi}~,\\
\chi_{ijmn}^{(2)} &=& \sum_{k=1}^\infty \left(\mp\right)^{k-1} k e^{k\f}
\nonumber\\
&&\times \int_0^{2\pi} \dd\psi \sin^i\psi \cos^j\psi  e^{k\a_\perp \cos\psi}\int_0^\infty\dd
\chi \sinh^m\chi \cosh^n \chi  e^{-k z_\perp\cosh\chi}~,
\eea
where the upper sign is for fermions and the bottom one is for bosons.
Using the integral representation of the modified Bessel functions of the 
first kind $I_n(x)$ and of the second kind $K_n(x)$ we can write
\bea
\chi_{ijmn}^{(1)} &=& \sum_{k=1}^\infty \left(\mp\right)^{k-1} e^{k\f}
i_{ij}(k\a_\perp) k_{mn}(kz_\perp)~,\\
\chi_{ijmn}^{(2)} &=& \sum_{k=1}^\infty \left(\mp\right)^{k-1} k e^{k\f}
i_{ij}(k\a_\perp) k_{mn}(kz_\perp)~,
\eea
where
\bea
i_{ij} &=&\sum_{l=0}^{[(1/2)d]} \left( \begin{array}{c}
       \displaystyle{[(1/2)d]} \\ 
       \displaystyle{ l}
       \end{array} \right) (2b+2l-1)!!  y^{-b-l} 2\pi I_{b+l+h}(y)~,\\
k_{mn} &=& \sum_{l=0}^{[(1/2)d]} \left( \begin{array}{c}
       \displaystyle{[(1/2)d]} \\ 
       \displaystyle{ l}
       \end{array} \right) (2b+2l-1)!!  y^{-b-l}  K_{b+l+h}(y)~,
\eea
and $[p]$ stands for the largest integer not exceeding $p$.
In $i_{ij}$, ($d = j$, $b=i/2$, $y = k\a_\perp$) while in $k_{mn}$, 
($d = n$, $b=m/2$, $y = k z_\perp$). Furthermore, $h$ is zero or one depending
on whether $d$ is even or odd respectively. Note also that for odd powers in
$\sin \psi$ the $i_{ij}$ vanishes and similarly for odd powers of $\sinh\chi$
the $k_{mn}$ vanishes.

In the case of pure longitudinal expansion $i=j=0$ and $\a_\perp=0$. 
Then $i_{ij}(k\a_\perp)=i_{00}(0)=2\pi$ and  
$\chi_{ijmn}$ reduces to
\bea
\chi_{00mn}^{(1)} &=& 2\pi \sum_{k=1}^\infty \left(\mp\right)^{k-1} e^{k\f}
k_{mn}(kz_\perp)~,\label{eq:chi1n}\\
\chi_{00mn}^{(2)} &=& 2\pi \sum_{k=1}^\infty \left(\mp\right)^{k-1} k e^{k\f}
k_{mn}(kz_\perp)~. \label{eq:chi2n}
\eea
The inclusive particle distribution and transverse energy  
at freeze-out are given by
\bea
{\dd N(\t_f) \over \dd y \dd^2 \tpt} &=& \int \t_f \dd \h \dd^2 \txt 
\mt \cosh \chi \, f(x,p)~,\\
{\dd \Et \over \dd y } &=& \int  \dd^2 \tpt \mt 
{\dd N(\t_f) \over \dd y \dd^2 \tpt} = \int\t_f \dd \h \dd^2\txt 
\int\dd^2\tpt \mts \cosh \chi \, f(x,p)~.
\eea
Using Eq.~(\ref{eq:noneqf}), for the distribution function $f(x,p)$, 
the particle spectra can be written as 
\be
{\dd N(\t_f) \over \dd y \dd^2 \tpt} = \int \t_f \dd \h \dd^2 \txt 
\mt \cosh \chi
\left\{f^{\eqq}(x,p)+f^{\eqq}(x,p)\Delta^{\eqq}(x,p)\f(x,p)\right\}~,
\label{eq:dndy}
\ee
where 
\be
f^{\eqq}(x,p) = A_0 {1\over \exp \{ m_{\bot}\cosh \chi/T \} -1}
~.\label{eq:1dfeq}
\ee
The first term in the curly brackets of Eq.~(\ref{eq:dndy}) is the equilibrium
contribution to the spectra while the second term is the non-equilibrium
contribution to the spectra.
Using the expression for $\f(x,p)$ and the integral representation $\chi_{ijmn}$
the particle distribution becomes
\bea
{\dd N(\t_f) \over \dd y \dd^2 \tpt} &=& A_0 {1\over 2}R_\perp^2\t_f 
 \times \biggl\{ \chi_{0001}^{(1)} + \left[ (\ca_0 -m^2 \ca_2)\chi_{0001}^{(2)} -\ca_1
 \mt\chi_{0002}^{(2)} +4\ca_2 \mts\chi_{0003}^{(2)}\right] \Pi \nonumber\\
 &&+\cc_0\left[\mts\chi_{0021}^{(2)}-{1\over
 2}\pts\chi_{0001}^{(2)}\right]\pi_s\biggr\}
\eea
where the first term which involves the $\chi_{ijmn}^{(1)}$ is the equilibrium
spectrum and the rest of the terms that involves $\chi_{ijmn}^{(2)}$ represent
the non-equilibrium contribution to the spectrum, i.e, they represent the
corrections to the particle spectra due to dissipation.
For dissipative corrections of particle spectra due to shear viscosity only  
we find, using Eqs.~(\ref{eq:chi1}) and (\ref{eq:chi2}),
\be
{\dd N(\t_f) \over \dd y \dd^2 \tpt}=A_0 \pi R_\perp^2\t_f \mt {1\over 4} {\pi_s \over \ve+p}
K_1(\zt)\biggl[\ats-\half \zts\biggl({K_3(\zt)\over K_1(\zt)}-1\biggr)\biggr]
\ee


In the case of a pure (1+1)-dimensional expansion in planar geometry,
the transverse dimension enters only as
a (constant) transverse area factor $A$.
The parametric integration over the hypersurface $\Sigma$ in Eq.~(\ref{eq:cf})
is performed by distinguishing space-like parts $\Sigma_s$
(with $z$ as integration variable) and time-like parts
$\Sigma_t$ (with $t$ as integration variable). This yields the
total momentum--space distribution
\begin{eqnarray} \label{hypercooper}
  \frac{{\rm d}N}{{\rm d} y p_{\bot}{\rm d}p_{\bot}} & = & 2\pi A
  \,m_{\bot}
  \left[
  \int_{\Sigma_s} {\rm d}z~
  [\cosh y - \sinh y (\partial t/\partial z)_{\Sigma}] f(x,p)
  \right. \nonumber \\
 &   & \left.
  +\int_{\Sigma_t} {\rm d} t~
  [\cosh y
  (\partial z/\partial t)_{\Sigma} - \sinh y] f(x,p)
  \right]~,\label{cp3}
\end{eqnarray}
where $(\partial t/\partial z)_{\Sigma}$  is the
(local) slope of the space-like hypersurface element. In the expression for the
distribution function Eq.~(\ref{eq:noneqf}) the equilibrium
distribution $f^{eq}$ is given by Eq.~(\ref{eq:1dfeq}) while $\f(x,p)$ is given
by Eq.~(\ref{eq:1dphi}). Note that in pure (1+1)-dimensional expansion we have
only one independent component of heat flux which we take to be $q^z =\g
\cq^z=\g q$ and one independent component of shear stress tensor which we take
to be $\pi^{zz} = \g^2 \t^{zz} = \g^2 \pi_s$. We simply write $q$ for the local
rest frame independent component of heat flux $\cq^z$ and $\pi_s$ for the local 
rest frame independent component of shear stress tensor $\t^{zz}$. From
\cite{AMI} we know that the other nonvanishing component of heat flux can be written as $q^0
= q^z v_z$ and the other nonvanishing components of the shear stress tensor are 
$\pi^{00} = \pi^{zz} v_z^2 = \pi_s \g^2 v_z^2$, 
$\pi^{0z}=\pi^{z0}=\pi^{zz} v_z = \pi_s \g^2 v_z$ and $\pi^{xx} = \pi^{yy} =
-\pi_s/2$.

We now apply Eq.~(\ref{cp3}) to the isothermal freeze--out
scenario. Along the isotherm, $T=T_f={\rm constant}$ and only $\h$
depends on position or time, respectively.
The rapidity distribution is obtained by integrating Eq.~(\ref{cp3})
over transverse momentum. For massless particles,
\begin{eqnarray}
  \frac{{\rm d}N}{{\rm d}y} & = & 2\pi A 
\left[  \int_{\Sigma_s} {\rm d}z~ {\cosh y -
    \sinh y (\partial t/\partial z)_{\Sigma} \over
     \cosh^3\chi(z)} \cf \right. \nonumber  \\
 &   & \left. + \int_{\Sigma_t} {\rm d} t~ {
    \cosh y (\partial z/\partial t)_{\Sigma} -\sinh y \over
    \cosh^3\chi(t)} \cf \right]~,\label{cp9}
\end{eqnarray}
where
\bea
\cf &=& {1\over 4\pi} \left[ \ci_{10} 
+ [\cj_{10}\ca_0 - \cj_{20}\ca_1+ 4\cj_{30}]\Pi
+[\cb_0\cj_{20}-\cb_1\cj_{30}] q \tanh \chi(z) \right.\nonumber\\
&&\left. +\cc_0\cj_{30} \pi_s\left(1-{3\over 2}{1\over \cosh^2
\chi(z)}\right) \right]~, \label{eq:cf1}
\eea
and the $\ci_{nk}$ and the $\cj_{nk}$ are given in Appendix
\ref{sec:ultra-rel} for single particle species. 
Analogously, the transverse momentum distribution is obtained integrating
over $y$,
\begin{eqnarray}
 \frac{{\rm d}N}{p_{\bot}{\rm d}p_{\bot}} & = &
  2\pi ~p_\bot ~ \left[
  \int_{\Sigma_s} {\rm d}z~\int_{-\infty}^{\infty}
   {\rm d}y~ [\cosh y -\sinh y (\partial t/\partial z)_{\Sigma}]\cf \right. \nonumber \\
 &   & \left. +  \int_{\Sigma_t} {\rm d}t~\int_{-\infty}^{\infty}
   {\rm d}y~[\cosh y (\partial z/\partial t)_{\Sigma} - \sinh y]\cf \right]~,
  \label{cp11}
\end{eqnarray}
where
\bea
\cf &=& A_0{1\over \exp\{(p_\bot/T_f) \cosh\chi(z)\} - a} \nonumber\\
&&+ A_0{\exp\{(p_\bot/T_f)\cosh \chi(z)\} \over 
[\exp\{(p_\bot/T_f) \cosh\chi(z)\} - a]^2} \f(x,p) \label{eq:cf2}
\eea
and 
\bea
\f(x,p) &=& \left[\ca_0 - \ca_1 p_\perp\cosh\chi(z) 
+4 p_\perp\cosh^2\chi(z)\ca_2\right]\Pi \nonumber\\
&&+ \left[\cb_0-\cb_1 p_\perp\cosh\chi(z)\right]p_\perp\sinh\chi(z)~q \nonumber\\
&&+\cc_0 \left[ p^2_\perp\sinh^2\chi(z) - {1\over 2}p_\perp^2\right]\pi_s ~.
\eea
In the isochronous freeze-out scenario the hypersurface $\Sigma$ has
no time-like part and the second term in Eq.~(\ref{cp3}) vanishes. Moreover,
also the second term in the numerator of the remaining term vanishes
due to $\partial t/\partial z = 0$ for this particular hypersurface.
The temperature along the hypersurface, however, is no longer constant.
Thus the rapidity distribution becomes
\begin{equation} \label{cp13}
  \frac{{\rm d}N}{{\rm d}y} = 2\pi^2 A
  \int_{-L-t_f}^{L+t_f} {\rm d}z 
  {\cosh y \over \cosh^3\chi(z)} \cf~,
\end{equation}
with $\cf(z)$ given by Eq.~(\ref{eq:cf1}) with $T_f$ now replaced by $T(z)$. 
The transverse momentum distribution is
\begin{equation} \label{cp14}
  \frac{{\rm d}N}{p_{\bot}{\rm d}p_{\bot}} = 2\pi A
  ~p_\bot~ \int_{-L-t_f}^{L+t_f} {\rm d}z~
  \int_{-\infty}^{\infty} {\rm d} y~ \cosh y ~\cf ~.
\end{equation}
where $\cf$ is given by Eq.~(\ref{eq:cf2}) with the temperature $T_f$ now
replaced by $T(z)$. The results of this section generalizes that of Ref.
\cite{BMGR} by including non-equilibrium (dissipative) effects in the freeze-out
prescription. The most recent applicatio of freeze-out prescription in the
presence of shear viscosity corrections has proven to be excellent in describing
the spectra of particles produced at RHIC \cite{BR}.
%
%
\section{Summary and conclusions}
\label{sec:summary}
%
%
We have presented a microscopic interpretation of dissipative fluid dynamics by
means of the 14-moment method (or 14-Field Theory: an effective kinetic theory
for relativistic non-equilibrium fluid dynamics). 
From the relativistic kinetic theory using Grad's 14 moment method one obtains 
relativistic causal fluid dynamics which is a field theory of the 14
fields of net charge density--particle flux and stress--energy--momentum. The
field equations are based on the conservation laws of net conserved charge,
energy-momentum and on a balance of fluxes. The equations satisfies the
principle of relativity, entropy principle and the requirement of hyperbolicity
(for causality). The resulting field equations contain only bulk viscosity,
shear viscosity and heat conductivity as unknown functions. All other
coefficients may be calculated from the equilibrium equations of state. 
The interface between the microscopic and microscopic in the causal dissipative
fluid dynamics is effected via  the standard transport coefficients.  The
comparison between the microscopic and the macroscopic descriptions provides
more information on the transport properties that govern the relaxation of
various dissipative processes.

From the kinetic theory approach the unknown phenomenological coefficients 
$\zeta,\, \k,\, \eta,\, \alpha_0,\,
\alpha_1,\, \beta_0,\, \beta_1,$ and $\beta_2$ can now be explicitly identified
from the knowledge of the collision term and the equation of state. 
We conclude that the 14-Field Theory of viscous, heat-conducting fluids is quite
explicit - provided we are given the thermal equation of state $p=p(\a,\b)$
-except for he coefficients $\cc_A$. These coefficients must be measured or
determined from the underlying theory/model which gave the equation of state.

We have also presented the relativistic thermodynamic integrals in a more general
formalism and also for different special cases which are relevant for
applications to nuclear collisions applications. These integrals are needed for
calculating the second-order relaxation coefficients. The results of the later
will be presented in detail somewhere.
\appendix
%

\section{ Relativistic Thermodynamic Integrals}
\label{sec:momint}

In relativistic kinetic theory one frequently encounters the moments of the 
distribution function describing local and global equilibrium.
The $n$th moment is defined by
\be
I^{\mu_1...\mu_n}(x) = \int \dd w~ p^{\mu_1}...p^{\mu_n} f(x,p)
\label{eq:in}~,
\ee
and the corresponding change of the $n$th moment due to variation of
$f(x,p)$ is
\be
\delta I^{\mu_1...\mu_n}(x)=\int \dd w~ p^{\mu_1}...p^{\mu_n} \delta
y \Delta(x,p) f(x,p)
\ee
where $y = \ln(f(x,p)/\Delta(x,p))$.  
The variations of moments involves the auxiliary moments
\be
J^{\mu_1...\mu_n}(x)=\int \dd w~ p^{\mu_1}...p^{\mu_n} \Delta(x,p) f(x,p)
\label{eq:jn}~.
\ee
These moments can be expanded in terms of symmetrized tensor product of $u^\mu$
and the metric tensor $g^{\mu\nu}$. We define for $2k\le
n$, the set of rank $n$ tensors
\be
\btu^{(2k} u^{n-2k)} \equiv {2^k! k!(n-2k)!\over n!} \sum_{\rm permutations}
\btu^{\mu_1\mu_2}...\btu^{\mu_{2k-1}\mu_{2k}} u^{\mu_{2k+1}}...u^{\mu_n}
\label{eq:ntensor}~,
\ee
where $\btu^{\mu\nu}$ is the projector $g^{\mu\nu}-u^\mu u^\nu$. The sum, which
runs over all distinct permutations of the tensor indices, has been divided by
the total number of those permutations. The number $k$ can take all integer
values between zero and the largest integer not exceeding $n/2$. The latter
value will be denoted as $[n/2]$.
These tensors possesses the orthogonality property
\be
\btu^{(2k}u^{n-2k)}\btu_{(2l}u_{n-2l)} = {(2k+1)!(n-2k)!\over n!}\delta_{ql} 
\label{eq:ortho}~.
\ee
which arises because of all different terms, differing only by the trivial permutations like
$\btu^{\mu\nu} \rightarrow \btu^{\nu\mu}$, $\btu^{\mu\a}\btu^{\nu\b} \rightarrow 
\btu^{\nu\b}\btu^{\mu\a}$, $u^\mu u^\nu \rightarrow u^\nu u^\mu$, only $2^k
k!(2k)!$ terms survived. If these trivial repetitions are lumped together the
expansion contains
\be
a_{nk} = \left(\begin{array}{c}
                n\\2k
		\end{array} \right) (2k-1)!!
\ee
terms. 
The double factorial notation stands for $(2k-1)!! =
1\cdot 3\cdot 5\cdot\cdot\cdot (2k-1)$ for $k=1,2,\cdots $ and the recursion
formula $(2k-1)!! = (2k+1)!!/(2k+1)$ extends the definition to negative integer
arguments. The moment integrals are then expanded in the form
\be
I^{\mu_1...\mu_n} = \sum^{[n/2]}_{k=0} a_{nk} \ci_{nk} \btu^{(2k}u^{n-2k)}~,~~~~~
J^{\mu_1...\mu_n} = \sum^{[n/2]}_{k=0} a_{nk} \cj_{nk} \btu^{(2k} u^{n-2k)} ~.
\label{eq:ijexpand}
\ee
The scalar coefficients $\ci_{nk}$ and $\cj_{nk}$, which depend on the parameters
$\a$ and $\b$ are found by contracting both sides of
Eqs.~(\ref{eq:in}) and (\ref{eq:jn}) with a tensor of the form (\ref{eq:ntensor}) and using the
orthogonality property (\ref{eq:ortho}). The results of such contractions are
\bea
\ci_{nk}(\alpha,\beta) &=& {A_0\over (2 k+1)!!} \int_0^\infty \dd w
\left[p^\alpha u_\alpha - p^\alpha p_\alpha\right]^k (p^\mu u_\mu)^{n-2
k}\nonumber\\
&&~~~~~~~~~~~~~~~~~~\times{1 \over e^{\beta p^\nu u_\nu
-\alpha} - a} \label{eq:Inq}~,\\
\cj_{nk}(\alpha,\beta) &=& {A_0\over (2 k+1)!!} \int_0^\infty \dd w
\left[p^\alpha u_\alpha - p^\alpha p_\alpha\right]^k (p^\mu u_\mu)^{n-2
k}\nonumber\\
&&~~~~~~~~~~~~~~~~~~\times{e^{\beta p^\nu u_\nu -\alpha} \over \left[e^{\beta p^\nu u_\nu
-\alpha} - a\right]^2} \label{eq:Jnq}~.
\eea
The above integrals are invariant scalars, therefore they can be evaluated in
any frame. We will evaluate them in the local rest frame. In this frame $u^\mu =
(1,0,0,0)$, so that $p^\mu p_\mu = p^2= m^2$ and $p^\m
u_\m=p^0=\sqrt{\vect{p}^2+m^2}$. We introduce spherical
polar coordinates $\dd^3 \tp =p^2 \,\dd p \,\dd\Omega$.
Integrating over $\dd\Omega$ and introducing new variables
\be
x = {p\over T}~,~~~~~z = {m\over T} ~,
\ee
the above thermodynamic integrals, Eqs.~(\ref{eq:Inq}) and (\ref{eq:Jnq}) 
may be written as 
\bea
\ci_{nk}(\phi,z) &=& {4\pi A_0  \over (2 k+1)!!} T^{n+2} \int_0^\infty dx\, x^{2(k+1)}
\,(x^2+z^2)^{(n-2 k -1)/2} \nonumber\\
&&~~~~~~~~~~~~~~~~~~~~\times {1\over e^{\sqrt{x^2+z^2}-\phi} -a}
\label{eq:inq}~,\\
\cj_{nk}(\phi,z) &=& {4\pi A_0  \over (2 k+1)!!} T^{n+2} \int_0^\infty
dx\, x^{2(k+1)}
\,(x^2+z^2)^{(n-2 k -1)/2}\nonumber\\
&&~~~~~~~~~~~~~~~~~~~~\times{e^{\sqrt{x^2+z^2}-\phi}\,\over 
\left\{e^{\sqrt{x^2+z^2}-\phi}
-a\right\}^2} \label{eq:jnq}~\\
&=& {1\over\b}\ci_{n-1,k-1} +{n-2k \over \b}\ci_{n-1,k} ~.\nonumber
\eea
The second line of Eq.~(\ref{eq:jnq}) is obtained by partial integration.
 
Derivatives of $\ci_{nk}(\phi,z)$ and $\cj_{nk}(\phi,z)$, Eqs.~(\ref{eq:inq})
and (\ref{eq:jnq}) can be expressed in terms of the $\cj_{nk}(\f,z)$:
\begin{eqnarray}
\dd \ci_{nk} &=& \cj_{nk} \dd \f - m^{-1}\cj_{n+1,k} \dd z ~\nonumber\\
           &=& \cj_{nk} \dd \a - \cj_{n+1,k} \dd \b ~,\label{eq:dInq}\\
z\, \dd \cj_{nk} &=& m\,[\cj_{n-1,k-1} +(n-2k)\cj_{n-1,k}]\dd \f - [\cj_{n,k-1}
+(n+1-2k)\cj_{n,k}] \dd z~, \nonumber\\
\b\, \dd \cj_{nk} &=& [\cj_{n-1,k-1} +(n-2k)\cj_{n-1,k}]\dd \a - [\cj_{n,k-1}
+(n+1-2k)\cj_{n,k}] \dd \b \label{eq:dJnq}~.
\end{eqnarray}
For the equation of state $p \equiv p(\eps,n)$ we need the derivative of $p$,
$\eps$ and $n$ with respect to ($\f$, $z$) or with respect to ($\a$, $\b$). 
From Eq.~(\ref{eq:dInq}) we have
\begin{eqnarray}
\dd n &\equiv& \dd \ci_{10} = \cj_{10} \dd\alpha -\cj_{20} \dd\beta \sks 
\dd\eps \equiv \dd \ci_{20} = \cj_{20} \dd\alpha
-\cj_{30} \dd\beta ~,\label{eq:dn}\\
\dd p &\equiv& \dd \ci_{21} = \cj_{21}\dd \alpha -\cj_{31} \dd\beta \sks 
\dd \ci_{31} = \cj_{31}\dd \alpha - \cj_{41} \dd \beta~.
\label{eq:dp}
\end{eqnarray}
From Eqs.~(\ref{eq:dn}) and (\ref{eq:dp}) the differentials of $\a(\eps,~n)$,
$\b(\eps,~n)$, and $p(\eps,~n)$ are written as
\bea
\dd \a &=&{1\over D_{20}}\left\{\cj_{30} \dd n - \cj_{20} \dd \eps\right\} \sks
\dd \b ={1\over D_{20}}\left\{\cj_{20} \dd n - \cj_{10} \dd \eps\right\}
\enspace, \\
\dd p &=&{1\over D_{20}}\left\{(\cj_{21}\cj_{30}-\cj_{31}\cj_{20}) \dd n 
                              - (\cj_{21}\cj_{20}-\cj_{31}\cj_{10}) \dd \eps\right\} \enspace.
\eea
The fundamental thermodynamic relation together with the first law of
thermodynamics can be expressed as
\be
nT \dd (s/n) = \dd\eps - h \dd n 
= (\cj_{20} -h \cj_{10}) \dd\alpha -(\cj_{30} - h \cj_{20})\dd\beta
~.\label{eq:tds}
\ee
From Eqs.~(\ref{eq:dn}), (\ref{eq:dp}) and (\ref{eq:tds}), the differentials of 
$\a(n,s/n)$ and $\b(n,s/n)$ can be written as 
\begin{eqnarray}
\dd \a &=&{1\over D_{20}}\left\{(\cj_{30} - h \cj_{20}) \dd n - \cj_{20}
nT\dd(s/n)\right\} ~,\label{eq:aSn}\\
\dd \b &=&{1\over D_{20}}\left\{(\cj_{20} - h \cj_{10}) \dd n - \cj_{10} nT\dd(s/n)
\right\}~,\label{eq:bSn}
\end{eqnarray}
and the differentials of $\a(\eps,s/n)$ and $\b(\eps,s/n)$ as 
\bea
\dd \a &=&{1\over h D_{20}}\left\{(\cj_{30} - h \cj_{20}) \dd \eps 
                - \cj_{30}nT\dd(s/n)\right\} \enspace,\\
\dd \b &=&{1\over h D_{20}}\left\{(\cj_{20} - h \cj_{10}) \dd \eps 
                  - \cj_{20} nT \dd(s/n)
\right\}\enspace.
\eea
Thus from Eqs.~(\ref{eq:dp}), (\ref{eq:aSn}) and (\ref{eq:bSn}) we obtain the
expression used in bulk pressure relaxation coefficients, namely
\be
\left.{\partial \ln \ci_{31}\over \partial \ln \ci_{10}}\right|_{s/n}
=\left.{\partial \ln \cj_{42}\over \partial \ln \ci_{21}}\right|_{s/n} 
= {\cj_{21}\over \cj_{42}}{1\over D_{20}}\left[\cj_{31}(\cj_{30} - h \cj_{20}) -
 \cj_{41}(\cj_{20}-h\cj_{10})\right] ~.
\ee
From the definitions of specific heats per net conserved charge
\begin{equation}
C_p = -\beta \left.{\PD(s/n) \over \PD\beta}\right|_p\,,\,\,C_V =
-\beta \left.{\PD(s/n)\over \PD\beta}\right|_n\,,\,\, \Gamma =C_p/C_V ~,
\end{equation}
one finds, with the help of Eqs.~(\ref{eq:dn}), (\ref{eq:dp}) and (\ref{eq:tds}),
\begin{eqnarray}
C_p - C_V &=&\beta^2(h \cj_{10} -\cj_{20})^2/(n \cj_{10})~,\\
\Gamma-1 &=& (h \cj_{10} - \cj_{20})^2 /D_{20}~.
\end{eqnarray}
From the definitions of the speed of sound and the compressibilities 
\be
c_s^2 = \left.{\PD p \over \PD \eps}\right|_{s/n} =
\left.{\PD p \over \PD \eps}\right|_n +{n\over \eps+p}
 \left.{\PD p \over \PD n}\right|_\eps ~, \;\;\;
\a_s={1\over n}\left.{\PD n\over \PD p}\right|_{s/n}~,\;\;\;
\k_T = {1\over n}\left.{\PD n\over \PD p}\right|_T~, 
\ee
we find, with the help of (\ref{eq:dn}), (\ref{eq:dp}) and (\ref{eq:tds}),
\be
c^2_s = {n^2 \over \b (\eps+p)\cj_{10}}\Gamma \sks 
\a_s  = {\b \cj_{10} \over n^2 \Gamma} \sks
\k_T  = {\b \cj_{10}\over n^2}
\ee
Thus for a given equation of state, i.e. pressure as a function of two
independent state variables, the other thermodynamic quantities can be obtained
as partial derivatives of the pressure with respect to $\a$ and $\b$ or to $\f$
and $z$. This means also that the second order coefficients are also
determined from the equation of state.

From the net charge and energy conservation equations 
(in the particle or Eckart frame)
\begin{eqnarray}
&&\dot{n} +n\theta = 0 ~,\\
&&\dot{\eps} +(\eps + p + \Pi)\theta +
(\btd_\mu q^\mu + \pi^{\mu\nu}\btd_\nu u_\mu) =
0~,
\end{eqnarray}
with the standard notations $\dot{A}\equiv u^\mu\PD_\mu A$ and $\theta\equiv \PD_\m u^\m$, we can solve for $\dot{\a}$ and $\dot{\b}$ by
simultaneously solving 
\begin{eqnarray}
&&\cj_{10}\dot{\a} - \cj_{20} \dot{\b} + \ci_{10}\theta = 0 ~,\\
&&\cj_{20} \dot{\a} - \cj_{30} \dot{\b} + (\ci_{20}+\ci_{21}+\Pi)\theta 
+(\btd_\mu q^\mu + \pi^{\mu\nu}\btd_\nu u^\mu) =0 ~,
\end{eqnarray}
where we have used Eq.~(\ref{eq:dn}).
We then find 
\begin{eqnarray}
\dot{\b} &=& -\frac{\cj_{20}\ci_{10} - \cj_{10}(\ci_{20}+\ci_{21}+\Pi)}{D_{20}} \theta
+\frac{\cj_{10}(\btd_\mu q^\mu + \pi^{\mu\nu}\btd_\nu u_\mu)}
{D_{20}} ~,\\
\dot{\a} &=& -\frac{\cj_{30}\ci_{10}
-\cj_{20}(\ci_{20}+\ci_{21}+\Pi)}{D_{20}}\theta
+\frac{\cj_{20}(\btd_\mu q^\mu + \pi^{\mu\nu}\btd_\nu
u_\mu)}{D_{20}} ~.
\end{eqnarray}

The $\ci_{nk}$ and $\cj_{nk}$ integrals, Eqs.~(\ref{eq:inq}) and
(\ref{eq:jnq}), can be written in terms of the more familiar functions
$\ck_n(\f,z),\,\cl_{n+1}(\f,z)$ defined, for $n \ge 0$, by (cf.
\cite{IS}) 
\begin{eqnarray}
\ck_n(\phi,z) &=& {1\over (2 n-1)!!}{1\over z^n} \int_0^\infty
{dx\, x^{2 n}\,(x^2+z^2)^{-1/2} \over e^{\sqrt{x^2+z^2}-\phi}-a}
\label{eq:kn}~,\\
\cl_{n+1}(\phi,z) &=& {1\over (2 n-1)!!}{1\over z^{n+1}} \int_0^\infty
{dx\, x^{2 n} \over e^{\sqrt{x^2+z^2}-\phi}-a} \label{eq:ln}~.
\end{eqnarray}
Partial integration of these functions leads to the following relations
\be
{\PD\over \PD \f}\ck_n = \cl_n ~, ~~~~~
{\PD\over \PD\f} \cl_{n+1} = 
-z^n {\PD \over \PD z}\left\{z^{-n} \ck_n\right\} =\ck_{n-1} + 
(2 n/z)\ck_n ~. \label{eq:prop}
\ee
In the special case of a Boltzmann gas $(a=0)$, these functions becomes the
Bessel functions of the second kind up to a factor of the chemical phase, 
\begin{equation}
\cl_n(\f,z) = \ck_n(\f,z) = e^\phi\ K_n(z)~,
\end{equation}
where $K_n(z)$ are the modified Bessel functions of the second kind.

By making use of the binomial expansion the integrals 
$\ci_{nk}$ and $\cj_{nk}$, Eqs.~(\ref{eq:inq}) and (\ref{eq:jnq}), 
can be expressed in terms of 
the $\ck_n$ and $\cl_{n+1}$, Eqs.~(\ref{eq:kn}) and (\ref{eq:ln}). 
For $n=0,1,...,$ and $k \le \frac{1}{2}n$, this yields
\be \label{eq:inq2}
\ci_{nk} = 4\pi A_0 m^{n+2} \sum_{l=0}^{\frac{1}{2}n-k} a_{nkl}~ z^{-(k+l+1)}
\left\{ \begin{array}{ll}
            \ck_{k+l+1} & \mbox{for even $n$} ~,\\
            \cl_{k+l+2} & \mbox{for odd $n$}~. 
	\end{array}
\right.
\ee
and differentiation of these results with respect to $\f$, (cf.
Eqs.~(\ref{eq:dInq}) and (\ref{eq:prop})),  then gives
\be \label{eq:jnq2}
\cj_{nk} = 4\pi A_0 m^{n+2}\sum_{l=0}^{\frac{1}{2}n-k} a_{nkl}~ z^{-(k+l+1)} 
\left\{ \begin{array}{ll}
           \cl_{k+l+1} & \mbox{for even $n$}\\
           \ck_{k+l} +{2(k+l+1)\over z}\ck_{k+l+1} & \mbox{for odd $n$}~. 
	\end{array}
\right. 
\ee
The numerical coefficients are 
\be
a_{nkl} = \frac{(2 k + 2 l + 1)!!}{(2 k+ 1)!!} 
\left( \begin{array}{c}
       \displaystyle{\frac{1}{2}n-k} \\ 
       \displaystyle{ l}
       \end{array} \right)  ~.
\ee
Note that in order to calculate the thermodynamic properties of
ultra-relativistic (massless) particles using Eqs.~(\ref{eq:inq2}) and
(\ref{eq:jnq2}) one has to take the limit $m\longrightarrow 0$ (hence
$z\longrightarrow 0$) of
Eqs.~(\ref{eq:kn}) and (\ref{eq:ln}). This is however not necessary if one uses
the integral representation Eqs.~(\ref{eq:inq}) and (\ref{eq:jnq}) since the
case for $m=0$ or $z=0$ is included. 
The integrals $\ci_{nk}$ and $\cj_{nk}$, Eqs.~(\ref{eq:inq2}) and
(\ref{eq:jnq2}), satisfy the recurrence relations
\be
{ \rm I}_{n+2,k} = m^2\,{\rm I}_{nk}+(2k+3) {\rm I}_{n+2,k+1} ~~~~~(2k\le n)  
\label{eq:recur} ~,
\ee
where ${\rm I}_{lm}$ stands for $\ci_{lm}$ or $\cj_{lm}$. This follows from
the contractions of Eq.~(\ref{eq:ijexpand}).

\section{Approximation to thermodynamic integrals}
\label{sec:limits}
%
\subsection{Nondegenerate gas}
\label{sec:non-degen}

If $\f \le z$ and $z \neq 0$, the integrals $\ci_{nk}$ and $\cj_{nk}$,
Eqs.~(\ref{eq:inq}) and (\ref{eq:jnq}),   
may be evaluated \cite{SC} by making the substitution $x=z\sinh \chi$
expressing the functions $(e^{z\cosh \chi} \pm 1)^{-1}$ as a geometric series
in $e^{z\cosh \chi}$ and integrating term by term. The integrals can be written
in terms of the Bessel functions with the help of 
\bea
{1\over e^{x-\f}\pm 1} = \sum_{k=1}^\infty (\mp)^{k-1} e^{-k(x-\f)}~, \\ 
{1\over [e^{x-\f}\pm 1]^2} = \sum_{k=1}^\infty (\mp)^{k-1} k e^{-k(x-\f)}~,
\eea
where the upper sign is for the fermions and the bottom one is for bosons. 
The integral representation of the Bessel functions of second
kind can be written as 
\be
\int_0^\infty \dd \chi \sinh^{2b}\chi \cosh^{d} \chi e^{-y\cosh \chi}
= \sum_{r=0}^{[(1/2)d]} \left( \begin{array}{c}
       \displaystyle{[(1/2)d]} \\ 
       \displaystyle{ r}
       \end{array} \right) (2b+2r-1)!!  y^{-b-r} K_{b+r+h}(y)~,
\ee
where $h$ is zero or one depending on whether $d$ is even or odd respectively.
Then the  integrals $\ci_{nk}$ and $\cj_{nk}$, Eqs.~(\ref{eq:inq}) and (\ref{eq:jnq}),  
 can be written as
\bea
\ci_{nk}^{\pm}(\f,z) &=& {4\pi A_0\over (2k+1)!!}|\gamma_Q|^{i_q}\gamma_p
                 \sum_{l=1}^\infty (\mp 1)^{l-1} 
		 \left[e^{l\f} + (-1)^n e^{-l\f}\right] \sum_{r=0}^{[(1/2)d]}
                   b_{nkr}~ y^{-b-r} K_{b+r+h}(y)~,\\
\cj_{nk}^{\pm}(\f,z) &=& {4\pi A_0 \over (2k+1)!!}|\gamma_Q|^{j_q} \gamma_p
               \sum_{l=1}^\infty (\mp 1)^{l-1} l 
	      \left[e^{l\f} + (-1)^{n+1}e^{-l\f}\right] \sum_{r=0}^{[(1/2)d]}
                 b_{nkr}~ y^{-b-r} K_{b+r+h}(y)~,
\eea
where we have abbreviated
\be
b_{nkr} \equiv\left( \begin{array}{c}
       \displaystyle{[(1/2)d]} \\ 
       \displaystyle{ r}
       \end{array} \right)(2b+2r-1)!! ~, 
~~~~~ b\equiv k+1~, ~~~~~ d\equiv n-2k~, ~~~~~ y\equiv lz ~.
\ee
The superscript notation $\pm$  on the $\ci_{nk}$ and $\cj_{nk}$ indicates
that the integrals are evaluated taking into account pair production. The factor
$\gamma_p$, which is 1 or 1/2 depending on whether the particle is charged or
neutral respectively, takes into account the contribution of neutral particles
in calculating thermodynamic properties of a gas. $\gamma_Q$ is the quantum
number for the conserved charge and $i_q$ is 1 or 0 depending on whether $n$ is
even or odd respectively while $j_q$ is 0 or 1 depending on whether $n$ is
even or odd respectively.
%
\subsection{Extremely degenerate Fermi gas}
\label{sec:degen}
This case is relevant to the study of the low temperature region in high energy
nuclear collisions and to the study of thermodynamic properties of neutron and
quark stars. 
If $\f \gg z$, the integrals $\ck_n$ and $\cl_{n+1}$, Eqs.~(\ref{eq:kn}) and
(\ref{eq:ln}), may be expressed in terms of
the Chandrasekhar-Sommerfeld asymptotic expansion. 
\bea
\ck_n(\f,z) &\sim& {1\over (2n-1)!!} {1\over z^n} \int_0^{x_0}
x^{2n}(x^2-z^2)^{-1/2}\dd x +2{\pi^2\over 12} {1\over (2n-3)!!} {1\over z^n}
(\f^2-z^2)^{(2n-3)/2}\f \nonumber\\
&&+2{7\pi^4\over 720}{1\over (2n-5)!!} {1\over
z^n}(\f^2-z^2)^{(2n-5)/2}\f\{3+(2n-5)(\f^2-z^2)^{-1}\f^2\}\nonumber\\
&&+ 2{31\pi^6\over 30240}{1\over (2n-7)!!}{1\over
z^n}(\f^2-z^2)^{(2n-7)/2}\f \nonumber\\
&&~~~\times\{15+10(2n-7)(\f^2-z^2)^{-1}\f^2
+(2n-7)(2n-9)(\f^2-z^2)^{-2}\f^4\}~,\\
\cl_{n+1}(\f,z) &\sim& {1\over (2n+1)!!} {1\over z^{n+1}} (\f^2-z^2)^{(2n+1)/2}
 \nonumber\\
&& +2{\pi^2\over 12} {1\over (2n-1)!!} {1\over z^{n+1}}
(\f^2-z^2)^{(2n-1)/2}\{1+(2n-1)(\f^2-z^2)^{-1}\f^2\}  \nonumber\\
&&+2{7\pi^4\over 720}{1\over (2n-3)!!} {1\over
z^{n+1}}(\f^2-z^2)^{(2n-3)/2} \nonumber\\
&&~~~\times\{3+6(2n-3)(\f^2-z^2)^{-1}\f^2
  + (2n-3)(2n-5)(\f^2-z^2)^{-2}\f^4\}\nonumber\\
&&+ 2{31\pi^6\over 30240}{1\over (2n-5)!!}{1\over
z^{n+1}}(\f^2-z^2)^{(2n-5)/2} \nonumber\\
&&~~~~~\times\{15 + 45(2n-5)(\f^2-z^2)^{-1}\f^2
+15(2n-5)(2n-7)(\f^2-z^2)^{-2}\f^4 \nonumber\\
&&~~~~~~~~~~~~~~~+(2n-5)(2n-7)(2n-9) (\f^2-z^2)^{-3}\f^6\}~,
\eea
with $x_0=\f$.
Changing the variable $x= z\sinh\chi$ one can evaluate the integral
\be
R_n \equiv {1\over z^n} \int_0^{x_0}
x^{2n}(x^2-z^2)^{-1/2}\dd x  = z^{2n} \int_0^{\chi_0} \sinh^{2n}\chi \dd \chi
~,
\ee
with the help of Eq.~(1.412.2) of Ref.~\cite{GR} and the properties of the hyperbolic functions.
Here we list the first few expressions for the $\ck_n$
\bea
\ck_0(\f,z) &\sim& \cosh^{-1}(\f/z) - {\pi^2\over6}{\f\over(\f^2-z^2)^{3/2}}
                   -{7\pi^4\over 120}{\f(2\f^2+3z^2) \over(\f^2-z^2)^{7/2}}
		 \nonumber\\
		 &&  -{31\pi^6\over 1008}{\f(8\f^4+40\f^2
		   z^2+15 z^4)\over(\f^2-z^2)^{11/2}}~,\\
z\ck_1(\f,z) &\sim& -{1\over2}z^2\cosh^{-1}(\f/z) + {1\over 2}\f(\f^2-z^2)^{1/2}
                    +{\pi^2\over6}{\f\over(\f^2-z^2)^{1/2}}
                    +{7\pi^4\over 120}{\f z^2\over(\f^2-z^2)^{5/2}}
		 \nonumber\\
		 && +{31\pi^6\over 1008}{\f z^2(4\f^4+3z^2)\over(\f^2-z^2)^{9/2}}
		   ~, \\
z^2\ck_2(\f,z) &\sim& {1\over 8} z^4 \cosh^{-1}(\f/z) 
                    + {1\over 24}\f(\f^2-z^2)^{1/2}(2\f^2-5z^2)
		    +{\pi^2\over 6}\f(\f^2-z^2)^{1/2}
		  \nonumber\\
		  &&  +{7\pi^4\over 360}{\f(2\f^2-3z^2) \over(\f^2-z^2)^{3/2}}
		    -{31\pi^6\over 1008}{\f z^4\over(\f^2-z^2)^{7/2}}~,\\
z^3\ck_3(\f,z) &\sim& -{1\over 48} z^6\cosh^{-1}(\f/z) 
                      +{1\over 720}\f(\f^2-z^2)^{1/2} (8\f^4-26\f^2z^2+3z^4)
		      +{\pi^2\over 18}\f(\f^2-z^2)^{3/2} 
		   \nonumber\\
		   &&   +{7\pi^4\over 360}{\f(4\f^2-3z^2) \over(\f^2-z^2)^{1/2}}
	+{31\pi^6\over 15120}{\f(8\f^4-20\f^2z^2 +15 z^4)
	\over(\f^2-z^2)^{5/2}}~,
\eea
and for the $\cl_{n+1}$
\bea
z\cl_1(\f,z) &\sim& (\f^2-z^2)^{1/2} - {\pi^2\over6}{z^2\over(\f^2-z^2)^{3/2}}
              -{7\pi^4\over 120}{z^2(4\f^2+z^2) \over(\f^2-z^2)^{7/2}}
	     \nonumber\\
	     && -{31\pi^6\over 336}{z^2(8\f^4+12\f^2z^2+z^4)
	     \over(\f^2-z^2)^{11/2}}~,\\
z^2\cl_2(\f,z) &\sim& {1\over 3} (\f^2-z^2)^{3/2} 
                     +{\pi^2\over6}{(2\f^2-z^2)\over(\f^2-z^2)^{1/2}}
		     +{7\pi^4\over 120}{z^4\over(\f^2-z^2)^{5/2}}
		  \nonumber\\
		  &&   +{31\pi^6\over
		  1008}{z^4(6\f^2+z^2)\over(\f^2-z^2)^{9/2}}~,\\
z^3\cl_3(\f,z) &\sim& {1\over 15}(\f^2-z^2)^{5/2} 
                     +{\pi^2\over 18}(\f^2-z^2)^{1/2}(4\f^2-z^2)
		     +{7\pi^4\over 360}{(8\f^4+3z^4)\over (\f^2-z^2)^{3/2}}
		    \nonumber\\
		    && -{31\pi^6\over 1008} {z^6\over (\f^2-z^2)^{7/2}}~,
\eea
that are needed for the calculation of the $\ci_{nk}$ and $\cj_{nk}$ 
used in the text.

For a massless Fermi gas the same calculation can be performed for the $\ci_{nk}$
and $\cj_{nk}$ directly using Eqs.~(\ref{eq:inq}) and
(\ref{eq:jnq})) to give 
\bea
\ci_{nk} &=& {4\pi A_0\over (2k+1)!!} T^{n+2}\Biggl[{\f^{n+2}\over n+2}\\
\nonumber 
        &&+ 2(n+1)!\left({\pi^2\over 12}{\f^n\over n!} + {7\pi^4 \over 720}{\f^{n-2}\over (n-2)!}
	                +{31\pi^6\over 30240}{\f^{n-4}\over (n-4)!} +
			\cdots\right)\Biggr]~,\\
\cj_{nk} &=& {4\pi A_0\over (2k+1)!!} T^{n+2}(n+1)\Biggl[{\f^{n+1}\over n+1}
\nonumber\\
       &&+  2 n!\left({\pi^2\over 12}{\f^{n-1}\over (n-1)!} 
	      + {7\pi^4 \over 720}{\f^{n-3}\over (n-3)!}
	                +{31\pi^6\over 30240}{\f^{n-5}\over (n-5)!} +
			\cdots\right)\Biggr]~.
\eea			
\subsection{Transition region}
\label{sec:transition}
The representation of the relativistic thermodynamic integrals as series of
modified Bessel functions is useful for computation in the non-degenerate case
and the representation in the degenerate region gets better for $\phi$ much
greater than $z$. A method suitable for the transition region $\phi \sim z$ is
needed for computing the values of the integrals in this region and for
numerical checking of known results.

The method adopted for the transition region is based on making the substitution
$x= t(2z+t^2)^{1/2}$ in the integrals. The integrals Eqs.~(\ref{eq:inq}) and
(\ref{eq:jnq}) becomes
\bea
\ci_{nk} &=& {4\pi A_0\over (2k+1)!!} T^{n+2}~
                  2 \int_0^\infty  {t^{2(k+1)}(z+t^2)^{n-2k}(2z+t^2)^{k+1/2} dt 
                                   \over e^{t^2+z -\f} - a}  \\
\cj_{nk} &=& {4\pi A_0\over (2k+1)!!} T^{n+2}~
                  2 \int_0^\infty  {t^{2(k+1)}(z+t^2)^{n-2k}(2z+t^2)^{k+1/2} dt 
                                   \over [e^{t^2+z -\f} - a]^2} 
\eea
while the function $\ck_n$ and $\cl_{n+1}$, Eqs.~(\ref{eq:kn}) and (\ref{eq:ln})
are written as 
\bea
\ck_{n} &=& {1\over (2n-1)!!} {1\over z^n}~ 2\int_0^\infty  {t^{2n}(2z+t^2)^{n-1/2} dt 
                                   \over e^{t^2+z -\f} - a}  \\
\cl_{n+1} &=& {1\over (2n-1)!!} {1\over z^n}~ 2\int_0^\infty  {t^{2n}(z+t^2)(2z+t^2)^{n-1/2} dt 
                                   \over e^{t^2+z -\f} - a} 
\eea
\section{Ultrarelativistic Thermodynamic Integrals}
\label{sec:ultra-rel}

\subsection{Single particle densities for massless particles}
\label{sec:part}

We want to study the thermodynamic properties of an ultrarelativistic gas.
Ultrarelativistic particles are characterized by vanishing rest mass. We start
by looking at the single particle densities.  For ultra-relativistic or
massless particles, $z=m/T=0$, the moment integrals,  Eqs.~(\ref{eq:inq}) and
(\ref{eq:jnq}) take the simple forms 
\bea
\ci_{nk} &=&{4\pi A_0 \over (2 k+1)!!} T^{n+2} \int_0^\infty dx\, x^{n+1} {1\over
e^{x-\f}-a} ~,\\
\cj_{nk} &=&{4\pi A_0 \over (2 k+1)!!} T^{n+2} \int_0^\infty dx\, x^{n+1}
{e^{x-\f}\over
\left[e^{x-\f}-a\right]^2} ~. 
\eea
For the fermions, when the equilibrium quantities are characterized by 
$\f\simeq 0$,
we can expand the the integrals in powers of $\f$. We have
\bea
\ci_{nk} &=&{4\pi A_0 \over (2 k+1)!!} T^{n+2} \int_0^\infty dx\, x^{n+1} 
\nonumber \\
&&\times \left[{1\over
e^{x-\f}+1} + \f {e^{x-\f}\over\left[e^{x-\f}+1\right]^2} 
+ {1\over 2}\f^2 (n+1) {x^n\over x^{n+1}}{e^{x-\f}\over\left[e^{x-\f}+1\right]^2}
\right] +\cdots~,\\
\cj_{nk} &=&{4\pi A_0 \over (2 k+1)!!} T^{n+2} (n+1)\int_0^\infty dx\, x^n 
\nonumber \\
&&\times \left[{1\over
e^{x-\f}+1} + \f {e^{x-\f}\over\left[e^{x-\f}+1\right]^2} 
+ {1\over 2}\f^2 n {x^{n-1}\over x^n} {e^{x-\f}\over\left[e^{x-\f}+1\right]^2}
\right] +\cdots~,
\eea

With the help of (cf. \cite{LL2})
\bea
\int_0^\infty \dd x\, x^{n+1} {1\over e^x+1} &=&
(1-2^{-(n+1)})\Gamma(n+2)\zeta(n+2) ~,\\
\int_0^\infty \dd x\, x^{n+1} {e^x\over(e^x+1)^2} &=&
(1-2^{-n})\Gamma(n+2)\zeta(n+1) ~,
\eea
where $\Gamma(x)$ is the Gamma function and $\zeta(x)$ is the Riemann's zeta
function, (Note that the $\zeta$ function with even argument is analytically
known),  we get, for a gas of quarks,
\bea
\ci_{nk} &=&{g_Q\over 2\pi^2}{1 \over (2 k+1)!!} T^{n+2}  
\Biggl[(1-2^{-(n+1)})\Gamma(n+2)\zeta(n+2) \nonumber\\
&&+ \f(1-2^{-n})\Gamma(n+2)\zeta(n+1)  
+ {1\over 2}\f^2 (n+1) (1-2^{-(n-1)})\Gamma(n+1)\zeta(n)
\Biggr] +\cdots~,\\
\cj_{nk} &=&{g_Q\over 2\pi^2}{1 \over (2 k+1)!!} T^{n+2}  (n+1)
\Biggl[(1-2^{-n})\Gamma(n+1)\zeta(n+1) \nonumber\\
&&+ \f(1-2^{-(n-1)})\Gamma(n+1)\zeta(n)  
+ {1\over 2}\f^2 n (1-2^{-(n-2)})\Gamma(n)\zeta(n-1)
\Biggr] +\cdots~,
\eea
with $g_Q$ the quark degeneracy. For the antiquarks we replace $\f$ by $-\f$.
For gluons, $\f=0$, and with the help of (cf. \cite{LL2})
\bea
\int_0^\infty \dd x\, x^{n+1} {1\over e^x-1} &=& \Gamma(n+2)\zeta(n+2) ~,\\
\int_0^\infty \dd x\, x^{n+1} {e^x\over(e^x-1)^2} &=& \Gamma(n+2)\zeta(n+1) ~,
\eea 
we have
\bea
\ci_{nk} &=&{g_G\over 2\pi^2}{1 \over (2 k+1)!!} T^{n+2}
                                \Gamma(n+2)\zeta(n+2)~,
\label{eq:ci0qg} \\
\cj_{nk}&=&{g_G\over 2\pi^2}{1 \over (2 k+1)!!} T^{n+2}
                               \Gamma(n+2)\zeta(n+1)~,  
\label{eq:cj0qg}
\eea
with $g_G$ the gluon degeneracy.

As an example, the number density, energy density and pressure of quarks  
are given by  
\bea
n_q &=& \ci_{10} = {g_Q\over 2\pi^2} T^3 \left[{3\over 2}\zeta(3) +\f
{\pi^2\over 6} \right] ~,\\ 
\eps_q &=& \ci_{20} = {g_Q\over 2\pi^2} T^4 \left[{7\over 4} {\pi^4\over
30} + \f {9\over 2}\zeta(3) + {1\over 2}\f^2 {\pi^2\over 2}\right]~,\\ 
p_q &=& \ci_{21} =  {1\over 3} \ci_{20}~.
\eea
and for the antiquarks one replaces $\f$ by $-\f$.
Similarly, the number density, energy density and pressure of gluons are
calculated using the boson integrals to give
\be
n_g = \ci_{10}=g_G {\zeta(3)\over \pi^2} T^3 \sks
\eps_g = \ci_{20}=g_G{\pi^2\over 30}  T^4 \sks
p_g = {1\over 3}\eps_g = \ci_{21}= {1\over 3} \ci_{20} ~.
\ee

We can also calculate the relaxation and coupling coefficients for
the quarks and gluons. For $\f=0$ one can immediately check that for massless particles the
quantity  $\Omega$ given by Eq.~(\ref{eq:om}) that comes in the relaxation and 
coupling coefficients for bulk
pressure vanishes identically. Thus those coefficients then diverges. That is,
$\Omega\Lra 0\,,\,\,\,\a_0\Lra \infty\,,\,\,\,\b_0\Lra \infty $. This have the
consequence that the bulk viscosity itself has to go to zero.
For the shear tensor relaxation coefficients, we have
for the quarks and gluons respectively 
\bea
\beta_2^q &=& {3\over 4}\times 31\times{8\over 15^2} {\zeta(6)\zeta(4)\over \zeta(5)
\zeta(5)} \,{1\over p} ~,\\
\beta_2^g &=& {3\over 4} {\zeta(6)\zeta(4)\over \zeta(5) \zeta(5)}\, {1\over p} ~.
\eea

For a classical Boltzmann gas in the ultra-relativistic limit $(z\ll 1)$
one gets the following expression for the shear relaxation coefficient
\be
\beta_2={3\over4}p^{-1}
~.
\ee
\subsection{A system of massless particles and antiparticles}
\label{sec:part-antipart}
%
We now  consider a gas of particles and antiparticles. For ultrarelativistic
Bose gas the chemical potential is zero and the contribution of bosons to the
thermodynamic properties of the gas is given by  Eqs.~(\ref{eq:ci0qg}) (with $n$
even, since for odd values of $n$ there are cancellations, e.g., net baryon
number vanishes) and (\ref{eq:cj0qg}) (with $n$ odd, since for even values of $n$
there are cancellations). The same argument goes for the ultrarelativistic
Fermi gas with vanishing chemical potential. But here we are considering a gas
at finite chemical potential.   
For ultrarelativistic or massless particles, $z=m/T=0$, the moment integrals, 
Eqs.~(\ref{eq:inq}) and (\ref{eq:jnq}) takes the simple forms 
\bea
\ci_{nk}^{\pm} &=&{4\pi A_0 \over (2 k+1)!!} T^{n+2} \int_0^\infty dx\, x^{n+1}\left\{ {1\over
e^{x-\f}+1} + (-1)^n {1\over e^{x+\f}+1} \right\} ~,\\
\cj_{nk}^{\pm} &=&{4\pi A_0 \over (2 k+1)!!} T^{n+2} \int_0^\infty dx\, x^{n+1}
\left\{ {e^{x-\f}\over \left[e^{x-\f}+1\right]^2} 
+(-1)^{n+1}{e^{x+\f}\over \left[e^{x+\f}+1\right]^2} \right\}~. 
\eea
These integrals can be evaluated by analytical means. We substitute $y^- =
x-\f$ in the first term and $y^+ = x+\f$ in the second term and then write the
integrals so that we can integrate from 0 to $\infty$.  Then we have 
\bea
\ci_{nk}^{\pm} &=&{4\pi A_0 \over (2 k+1)!!} T^{n+2} \Biggl[ \int_0^\infty dy^+\, 
{(y^+ +\f)^{n+1} \over e^{y^+}+1} 
+ (-1)^n \int_0^\infty dy^-\,{(y^- -\f)^{n+1} \over e^{y^-}+1} \nonumber\\
&&+\int_{-\f}^0 dy^+\, {(y^+ +\f)^{n+1}\over e^{y^+}+1} -(-1)^n \int_0^\f dy^-\,
{(y^- - \f)^{n+1} \over e^{y^-}+1} \Biggr] ~,\\
\cj_{nk}^{\pm} &=&{4\pi A_0 \over (2 k+1)!!} T^{n+2} (n+1)\Biggl[ \int_0^\infty dy^+\, 
{(y^+ +\f)^n \over e^{y^+}+1} 
+ (-1)^{n+1} \int_0^\infty dy^-\,{(y^- -\f)^n \over e^{y^-}+1} \nonumber\\
&&+\int_{-\f}^0 dy^+\, {(y^+ +\f)^n\over e^{y^+}+1} -(-1)^{n+1} \int_0^\f dy^-\,
{(y^- - \f)^n \over e^{y^-}+1} \Biggr]~. 
\eea
For $\cj_{nk}$ we first perform partial integration. 
The first two integrals can be directly combined and the last two after the
substitution $y^- = -y^+$. We then have 
\bea
\ci_{nk}^{\pm} &=&{4\pi A_0 \over (2 k+1)!!} T^{n+2} \left[\int_0^\infty dx\, 
{(x +\f)^{n+1} + (-1)^n (x -\f)^{n+1} \over e^x+1 }
+\int_0^\f dz\, z^{n+1}\right]~,\\
\cj_{nk}^{\pm} &=&{4\pi A_0 \over (2 k+1)!!} T^{n+2} (n+1)\left[\int_0^\infty dx\, 
{(x +\f)^n + (-1)^{n+1} (x -\f)^n \over e^x+1 }
+\int_0^\f dz\, z^n \right]~.
\eea
In the last two integrals we noted that $(e^x+1)^{-1} +(e^{-x}+1)^{-1}=1$ and we
have substituted $z= x+\f$. The two power functions in the numerator are then
expanded according to the binomial expansion. 
All terms of even $j$ drops out:
\bea
\ci_{nk}^{\pm} &=&{4\pi A_0 \over (2 k+1)!!} T^{n+2} \nonumber\\
&&\times\left[
2\sum_{j=1,3,5,\cdots \leq(n+1)}\left( \begin{array}{c}
       \displaystyle{n+1} \\ 
       \displaystyle{ j}
       \end{array} \right)\f^{n+1-j} \int_0^\infty dx\, {x^j \over e^x+1}
+\int_0^\f dz\, z^{n+1} \right]~,\\
\cj_{nk}^{\pm} &=&{4\pi A_0 \over (2 k+1)!!} T^{n+2} \nonumber\\
&&\times(n+1)\left[
2 \sum_{j=1,3,5,\cdots \leq n}\left( \begin{array}{c}
       \displaystyle{n} \\ 
       \displaystyle{ j}
       \end{array} \right)\f^{n-j} \int_0^\infty dx\, {x^j \over e^x+1}
+\int_0^\f dz\, z^{n}\right]~.
\eea
The above integrals can be brought to
their final analytic form by making use of
\be
\int_0^\infty dx\, {x^j \over e^x+1} = \Gamma(j+1)\zeta(j+1) \left(1-{1\over
2^j}\right) ~.
\ee
 Thus we finally have
\bea
\ci_{nk}^{\pm} &=&{4\pi A_0 \over (2 k+1)!!} T^{n+2} \nonumber\\
&&\times\left[
2\sum_{j=1,3,5,\cdots\leq(n+1)}\left( \begin{array}{c}
       \displaystyle{n+1} \\ 
       \displaystyle{ j}
       \end{array} \right)\f^{n+1-j} 
       \Gamma(j+1)\zeta(j+1) \left(1-{1\over 2^j}\right) + {\f^{n+2}\over n+2}
        \right] \label{eq:ciana}\\
\cj_{nk}^{\pm} &=&{4\pi A_0 \over (2 k+1)!!} T^{n+2} \nonumber\\
&&\times(n+1)\left[
2 \sum_{j=1,3,5,\cdots\leq n}\left( \begin{array}{c}
       \displaystyle{n} \\ 
       \displaystyle{ j}
       \end{array} \right)\f^{n-j}
       \Gamma(j+1)\zeta(j+1) \left(1-{1\over 2^j}\right) + {\f^{n+1}\over n+1} 
       \right] \label{eq:cjana}
\eea
For the
total particle number densities or single particle species densities separately
one can no longer use the above formulation. One then uses the results from the
previous section.

To this end we give here the results of the ultrarelativistic thermodynamic
integrals $\ci^\pm_{nk}$ and $\cj^\pm_{nk}$ for QGP. We present only those that
are used in the text. We treat quarks and gluons as ultrarelativistic Fermi or Bose
gases, respectively. Then from Eqs.~(\ref{eq:ciana}) and (\ref{eq:cjana}) we
have , for a gas of quarks and gluons,
\bea
\ci_{10}^{\pm} &=& {g_Q\over 6}T^3\left[\f +{\f^3\over \pi^2}\right]~,\\
\ci_{20}^{\pm} &=& T^4\left[\left(g_G + {7\over 4} g_Q\right){\pi^2\over 30} 
                             +g_Q\left({\f^2\over 4} + {\f^4\over
			     8\pi^2}\right)\right] \sks
\ci_{21}^{\pm} = {I_{20}^{\pm}\over 3}~,\\
\ci_{30}^{\pm} &=& g_Q T^5\left[{7\pi^2 \over 30}\f + {\f^3\over 3} + {\f^5\over
                10\pi^2}\right] \sks
\ci_{31}^{\pm} = {\ci_{30}^{\pm}\over 3}~,\\
\ci_{40}^{\pm} &=& T^6\left[\left(g_G+{31\over 16}g_Q\right){4\pi^4\over 3\times 21}
+{g_Q\over 12}\left(7\pi^2\f^2 +5\f^4 +{\f^6\over \pi^2}\right)\right]~,\\
\ci_{41}^{\pm} &=& {\ci_{40}\over 3} \sks \ci_{42}^{\pm} = {\ci_{40}^{\pm}\over 15}~,\\
\cj_{10}^{\pm} &=& {1\over 2} T^3\left[{1\over 3}(g_G+g_Q) + {g_Q\over
               \pi^2}\f^2\right]~,\\
\cj_{20}^{\pm} &=& {g_Q\over 2} T^4\left[\f + {\f^3\over \pi^2}\right] \sks
\cj_{21}^{\pm} = {\cj_{20}^{\pm}\over 3}~,\\
\cj_{30}^{\pm} &=& T^5\left[\left(g_G+{7\over 4}g_Q\right){2\pi^2\over 15}
                      +g_Q\left(\f^2+{\f^4\over 2\pi^2}\right)\right] \sks
\cj_{31}^{\pm} = {\cj_{30}^{\pm}\over 3}~,\\
\cj_{40}^{\pm} &=& g_Q T^6\left[{7\pi^2\over 6}\f + {5\over 3} \f^3 +{\f^5\over
             2\pi^2}\right] \sks
\cj_{41}^{\pm} = {\cj_{40}\over 3} \sks \cj_{42}^{\pm} = {\cj_{40}^{\pm}\over 15}~,\\
\cj_{50}^{\pm} &=& T^7\left[\left(g_G+{31\over 16}g_Q\right){8\pi^4\over 21}
+{g_Q\over 2}\left(7\pi^2\f^2 +5\f^4 +{\f^6\over \pi^2}\right)\right]~,\\
\cj_{51}^{\pm} &=& {\cj_{50}^{\pm}\over 3} \sks\cj_{52}^{\pm} = {\cj_{50}^{\pm}\over 15}~,
\eea
where $g_G=N_s(N_c^2-1)$ and $g_Q=N_s N_c N_f$, (with $N_s$ the number of spin 
projections, $N_c$ the number of color charges and $N_f$ the number of quark
flavours) are the quark and gluon 
degeneracies respectively. Note that the net baryon charge is $n=\ci^\pm_{10}$ while
the total energy density and the total pressure are given by $\eps=\ci^\pm_{20}$ and
$p=\ci^\pm_{21}$.
%
\section{Relativistic thermodynamic integrals at zero temperature, T=0}
\label{sec:T=0}
%
At $T=0$ the mean occupation number is in good approximation described by the
step function and the relativistic thermodynamic integrals Eqs.~(\ref{eq:Inq}) 
and (\ref{eq:Jnq}) becomes
\bea
\ci_{nk} &=& {4\pi A_0\over (2k+1)!!}\int_0^{p_f} (p^2+m^2)^{(n-2k-1)/2} p^{2(k+1)}
\dd p \label{eq:cink0}\\
\cj_{nk} &=& 0
\eea
The $\cj_{nk}$ integrals vanishes at $T=0$ because they are the product of
temperature and the delta function or the $\ci_{nk}$ integrals.
To calculate the integrals Eq.~(\ref{eq:cink0}) we make the substitution $p =
m\sinh \c$. Then we have $\dd p = m\cosh\c\, \dd \c$. If we set $p_f = m\sinh
\c_f$ then the integrals Eq.~(\ref{eq:cink0}) becomes
\be
\ci_{nk} = {4\pi A_0\over (2k+1)!!} m^{n+2} \int_0^{\c_f} \cosh^{n-2k} \c
\sinh^{2(k+1)} \c \,\dd \c \label{eq:ciT0}
\ee
For massless particles Eq.~(\ref{eq:cink0}) becomes
\bea
\ci_{nk} &=& {4\pi A_0\over (2k+1)!!}  \int_0^{p_f} p^{n+1} \dd p \nonumber \\
         &=& {4\pi A_0\over (2k+1)!!} {p_f^{n+2} \over n+2}
\eea

The integrals Eq.~(\ref{eq:ciT0}) are evaluated with the help of Eq.~(2.413) of
Ref.~\cite{GR} and the properties of the hyperbolic functions. Here we list the results which are used in the text:
\bea
\ci_{10} &=& 4\pi A_0 m^3\left[{1\over 3} \sinh^3 \c_f\right]\\
\ci_{20} &=& 4\pi A_0 m^4\left[-{\c_f\over 8}+{1\over 32}\sinh 4\c_f\right]\\
\ci_{21} &=& {4\pi A_0\over 3}m^4 \left[{3\over 8}\c_f -{3\over 8}\sinh \c_f \cosh\c_f 
                             +{1\over 4} \sinh^3\c_f \cosh\c_f\right]\\
\ci_{30} &=& 4\pi A_0 m^5\left[{1\over 5}\left(\cosh^2\c_f+{2\over3}\right)
                            \sinh^3\c_f\right]\\
\ci_{31} &=& {4\pi A_0\over 3} m^5\left[{1\over 5} \sinh^5 \c_f\right]
\eea
The Fermi momentum $p_f$ can be determined from the net charge density
\be
p_f = \left({3\over 4\pi A_0} \ci_{10}\right)^{1/3} \\
\ee
from which $\c_f =  \sinh^{-1} \left(p_f/ m\right)$


\acknowledgments
I would like to thank Rory Adams, Tomoi Koide, Pasi Huovinen and Etele Molnar 
for reading  the manuscript and for valuable comments.


%
%
\newpage

\begin{thebibliography}{99}  
%
%
%
\bibitem{AMI} A. Muronga, nucl-th/0611090

\bibitem{Chapman} S. Chapman and T.G. Cowling, {\it The Mathematical Theory of
Non-Uniform Gases}, Third Edition (Cambridge University Press, Cambridge, 1970).

\bibitem{HG} H. Grad, Commun. Pure Appl. Math. {\bf 2} (1949) 331.

\bibitem{CE} C. Eckart, Phys. Rev. {\bf 58} (1940) 919.

\bibitem{LL}
L.D.\ Landau and E.M.\ Lifshitz, {\it Fluid Mechanics} (Pergamon,
New York, 1959)

\bibitem{IM} I. M\"uller,  Z. Phys. {\bf 198} (1967) 329.

\bibitem{IS}
W.\ Israel, Ann.\ Phys.\ (N.Y.) {\bf 100} (1976) 310;
J.M.\ Stewart, Proc.\ Roy.\ Soc. A {\bf 357} (1977) 59;
W.\ Israel and J.M.\ Stewart, Ann.\ Phys.\ (N.Y.) {\bf 118} (1979) 341.

\bibitem{deGroot}
S.R.\ deGroot, W.A.\ van Leeuwen, and Ch.G.\ van Weert,
{\it Relativistic Kinetic Theory} (North-Holland, Amsterdam, 1980)

\bibitem{AMY} P. Arnold, G.D. Moore and L.G. Yaffe, JHEP {\bf 0112} (2001)
009.

\bibitem{AM} A. Muronga, Phys. Rev. {\bf C} (2004).

\bibitem{LMR}
I.-S Liu, I. M\"uller and T. Ruggeri, Ann. Phys. {\bf 169} (1986) 191.
%
\bibitem{GL}
R. Geroch, L. Lindblom, Phys. Rev. D {\bf 41} (1990) 1855.

\bibitem{walecka} See e.g.\ B.D.\ Serot, J.D.\ Walecka,
{\it The Relativistic Nuclear Many--Body Problem} in:
Adv.\ Nucl.\ Phys.\ 16 (1986) 1 (eds.\ J.W.\ Negele and E.\ Vogt, Plenum Press,
New York).

\bibitem{DHR} D. H. Rischke, Y. P\"urs\"un, and J. A. Maruhn, Nucl. Phys. {\bf
A595}, 383 (1995), Erratum-ibid. a596, 717 (1996).

\bibitem{migdhr} M.I.\ Gorenstein, D.H.\ Rischke, H.\ St\"ocker, W.\ Greiner,
J.\ Phys.\ G 19 (1993) L69.

\bibitem{gori} K.A.\ Bugaev, M.I.\ Gorenstein, Z.\ Phys.\ C 43
(1989) 261.

\bibitem{MIT} A.\ Chodos, R.L.\ Jaffe, K.\ Johnson, C.B.\ Thorn,
V.F.\ Weisskopf, Phys.\ Rev.\ D 9 (1974) 3471.

\bibitem{Landau} L.D.\ Landau, Izv.\ Akd.\ Nauk SSSR 17 (1953) 51, in:
{\it Collected papers of L.D.\ Landau} (ed.\ D.\ Ter--Haar, Pergamon, Oxford,
1965), p.\ 569--585, \\
L.D.\ Landau, S.Z.\ Belenkii, Uspekhi Fiz.\ Nauk 56 (1955) 309, ibid., p.\
665--700.

\bibitem{CF} F.\ Cooper, G.\ Frye, Phys.\ Rev.\ D 10 (1974) 186,\\
F.\ Cooper, G.\ Frye, E.\ Schonberg, Phys.\ Rev.\ D 11 (1975) 192.

\bibitem{Bj}
J.D. Bjorken, Phys. Rev. D {\bf 27} (1983) 140.

\bibitem{MR} A. Muronga and D.H. Rischke, nucl-th/0407114.
%
\bibitem{BMGR} S. Bernard, J.A. Maruhn, W. Greiner and D.H. Rischke, Nucl. Phys. {\bf A605} ,
566 (1996) 
%
\bibitem{BR}  R. Baier and  P. Romatschke, nucl-th/0610108
%
\bibitem{SC} S. \ Chandrasekhar,  {\it Introduction to the  study of
staellar structure}, chap. 10 (Chicago University Press 1939, reprint Dover, New
York, 1957);\\  
R.F. Tooper, Astrophys. J. {\bf 156} (1969) 1075.

\bibitem{GR} I.S. Gradsteyn and I.M. Ryzhik, {\it Tables of Integrals, Series, and
Products} (Academic Press, San Diego, 1980).

\bibitem{LL2}
L.D.\ Landau and E.M.\ Lifshitz, {\it Statistical Physics} (Pergamon Press,
Oxford, Third Edition, 1980)
%
\bibitem{MA} M. Abramowitz and I. A. Stegun, {\it Handbook of Mathematical
Functions} (Dover publications, New York, 1970).
%

\end{thebibliography}
\end{document}